\documentclass[twocolumn]{autart}    % Enable this line and disable the
\usepackage{tabularx}
\usepackage{amssymb}
\usepackage{graphicx}
\usepackage{amsmath}
\usepackage{mathabx}
\usepackage{mathrsfs}
\usepackage{verbatim}
\usepackage{flexisym}
\usepackage{setspace}
\usepackage{arydshln}
\usepackage{stackengine}
\usepackage{color}
\def\delequal{\mathrel{\ensurestackMath{\stackon[1pt]{=}{\scriptstyle\Delta}}}}
\usepackage{accents}
\newlength{\dhatheight}
\newcommand{\doublehat}[1]{%
    \settoheight{\dhatheight}{\ensuremath{\hat{#1}}}%
    \addtolength{\dhatheight}{-0.35ex}%
    \hat{\vphantom{\rule{1pt}{\dhatheight}}%
    \smash{\hat{#1}}}}

\begin{document}

\begin{frontmatter}

\title{A Data-driven Approach to Actuator and Sensor Fault Detection, Isolation and Estimation in Discrete-Time Linear Systems} % Title, preferably not more

\vspace{-0.5cm}
\author[conc]{E. Naderi and K. Khorasani} \ead{kash@ece.concordia.ca}    % Add the
% \author[conc]{K. Khorasani}\ead{kash@ece.concordia.ca}

\address[conc]{Department of Electrical and Computer Engineering, Concordia University, Montreal, Quebec, Canada}  % Please supply

                             % keyword list or with the
                                          % help of the Automatica
                                          % keyword wizard
\renewcommand{\baselinestretch}{0.8}

\vspace{-0.5cm}
\begin{abstract}                          % Abstract of not more than 200 words.
In this work, we propose explicit state-space based fault detection, isolation and estimation filters that are data-driven and are directly identified and constructed from only the system input-output (I/O) measurements and through estimating the system Markov parameters. The proposed methodology does not involve a reduction step and does not require identification of the system extended observability matrix or its left null space. The performance of our proposed filters is directly connected to and linearly dependent on the errors in the Markov parameters identification process. {The estimation filters operate with a subset of the system I/O data that is selected by the designer. It is shown that the proposed filters provide asymptotically unbiased estimates by invoking low order filters as long as the selected subsystem has a stable inverse. We have derived the estimation error dynamics in terms of  the Markov parameters identification errors and have shown that they can be directly synthesized from the healthy system I/O data. Consequently, the estimation errors can be effectively compensated for. Finally, we have provided several illustrative case study simulations that demonstrate and confirm the merits of our proposed schemes as compared to methodologies that are available in the literature.}
\end{abstract}

\end{frontmatter}

\renewcommand{\baselinestretch}{0.8}
\section{Introduction}

Since the concept of autonomous fault diagnosis, also known as fault detection and isolation (FDI), has been introduced by Beard \cite{Beard}, it has received enormous an attraction in the literature. Some excellent surveys have been published that summarize  the extensive literature on fault diagnosis of dynamical systems\cite{Gao2015,Gao2}. Two main fault diagnosis categories are model-based and data-driven approaches. The available model-based approaches are reviewed and described in \cite{hwang2010survey}. 

As engineering systems evolve, it is less likely that engineers have a detailed and accurate mathematical description of the dynamical systems they work with. On the other hand, advances in sensing and data acquisition systems can provide a large volume of raw data for most engineering applications. Consequently, one can find a trend towards data-driven based approaches in many disciplines and problems, including fault diagnosis.

The term `data-driven' covers a wide range of techniques in the literature. Some of the most important strategies are neural networks \cite{SinaTayaraniBathaie201522},   fuzzy logic \cite{RodríguezRamos2016}, and hybrid approaches \cite{SobhaniTehrani201412}. In addition to artificial intelligence based methods, some efforts have been made that are aimed at extending the rich model-based fault diagnosis techniques to  data-driven based approaches.\\ 

A trivial solution will be the one where one can first identify a mathematical dynamical model of the system from the available data, and then by using the resulting explicit model one then implements and designs conventional model-based schemes. However, this approach suffers from the subsequent errors that are introduced in the system identification process and that may ultimately aggravate the FDI scheme design process errors which can result in a totally unreliable fault diagnosis scheme.

In recent years, a new paradigm has emerged in the literature that aims at direct and explicit construction of the FDI schemes from the available system input-output (I/O) data \cite{ding2009subspace,dong2009subspace2,wang2011subspace}. Subspace-based data-driven fault detection and isolation methods  \cite{ding2014data,dong2009subspace} represent as one of the main approaches that have been reviewed in \cite{ding2014data}. These methods are developed based on identifying the left null space of the system extended observability matrix using the I/O data. An estimate of the system order and an orthogonal basis for the system extended observability matrix - or its left null space - are obtained via the SVD decomposition of a particular data matrix that is constructed from the system I/O data. This process is known as the \underline{reduction step}. \\

Essentially, in the reduction step  it is assumed that the number of the first set of significantly nonzero singular values and the associated directions provide an estimate of the system order and a basis for the extended observability matrix.  However,  in most cases, this process leads to erroneous results due to the fact that the truncation point for neglecting small singular values, as being insignificant, is not obvious a trivial and is subjective and problem dependent. \\

Consequently, an erroneous system order and basis for the extended observability matrix - or its left null space - can be obtained. This error  manifests itself in the fault diagnosis scheme performance in a nonlinear manner. In other words, the performance representation of the FDI scheme is not  a linear function of the gap between the estimated system order and the system extended observability matrix and the actual ones.   Due to these drawbacks, other works that have appeared in the literature are mainly concerned with only the fault estimation problem in which the main objective is to eliminate and remove the above reduction step. 

{Dong and his colleagues \cite{dong2012robust} have developed a fault detection scheme that can be directly synthesized from the system I/O data without involving the reduction step. The detection filter is in fact a high order FIR filter parameterized by the system Markov parameters. The extension of this work to the fault isolation task is not trivial and straightforward.  It can be performed by obtaining a projection vector that is computed through the SVD decomposition of a transfer matrix parameterized by the Markov parameters estimation errors \cite{Dong2012Isolation}. However, the Markov parameters estimation errors are not generally available. Therefore, the authors in \cite{Dong2012Isolation} have managed to synthesize this matrix from the I/O data. The order of the isolation filters can be as large as 30. Dong and Verhaegen \cite{dong2012identification} used the same strategy for direct construction of the fault estimation filter. The underlying assumption is that the system should have a stable inverse. It will be asymptotically unbiased if the FIR filter order tends to \textit{infinity}. \\}

{Wan and his colleagues \cite{wan2015data} have reasoned in their recent work  that the method of \cite{dong2012identification} cannot be applied to certain open-loop systems. Moreover, it does not compensate for the estimation errors. Consequently, Wan and his colleagues have proposed offline and online algorithms for compensating for the estimation errors. Yet, it suffers from \textit{two} major drawbacks. First, the estimation is asymptotically unbiased if the filter order tends to infinity. Second, the computational time per sample for the online optimization algorithm -  which is the one that yields an almost unbiased results among the others proposed - is significantly high as compared to the offline methods in \cite{wan2015data}. \\}

{In this work, to overcome the above drawbacks and limitations, we have proposed fault detection, isolation and estimation filters that are constructed directly in the state-space representation form from and using \underline{only} the available system I/O data. Our proposed schemes only require identification of the system Markov parameters that are achieved by using conventional methods, such as  correlation analysis \cite{ljung1998system} or subspace methods  \cite{van1995unifying,katayama2006subspace,huang2008dynamic,chiuso2007relation} from the healthy I/O data. \\}

{Our method does \textit{not} involve the reduction step or equivalent forms of the extended observability matrix. Therefore, the estimation error is linearly dependent on the Markov parameter estimation errors. This step is already addressed in the literature as reviewed above. However, it turns out that our state-space based approach can address several important difficulties that are associated with the currently available works in the literature. First, our proposed identification and isolation filters are conveniently configured for the isolation task of both single \textit{as well as} concurrent faults through constructing filter banks. \\}

{An important feature of our proposed state-space based method is  that  estimation will be achieved asymptotically unbiased by a filter order \textit{as low as} the maximum of the system relative degree and the system observability index. Both of these parameters are bounded by the system order. Moreover, it does not necessarily require the condition of having an entire stable inverse system. The flexibility of our proposed scheme allows arbitrary selection of the subsystems for achieving the fault isolation or for performing the fault estimation tasks.\\ }

{In other words, one can select a different subsystem if an actuator fault estimation is blocked due to unstable inversion of a specific subsystem. Finally, the state-space based approach allows one to implement a simple and yet effective procedure for compensating for the estimation errors. Towards this end, in this work we derive the estimation error dynamics and show that it can be directly identified from the healthy system I/O data. We will demonstrate through comprehensive simulation studies the effectiveness of our proposed error compensating producer. \\}

{The \underline{contributions} of this paper can now be summarized as follows: 
\begin{enumerate}
\item A general fault detection and isolation filter for both actuator and sensor faults is developed and directly constructed from only the available system I/O data in the state-space form in a manner that does not involve a reduction step. Moreover,  our approach does not require an \textit{a priori} knowledge of the system order. The proposed fault detection and isolation filters can be conveniently configured for both single and concurrent fault detection and isolation tasks by using a subset of the I/O data.
\item A fault estimation scheme for both actuator and sensor faults (single and concurrent) is developed and directly constructed from the available system I/O data in the state-space form in a manner that does not involve a reduction step. The proposed estimation filter is asymptotically unbiased having an order as small as the maximum of the observability index and the system relative degree.
\item A new offline procedure for \textit{tuning} the estimation filters are proposed to compensate for errors that are caused by the Markov parameters estimation uncertainties.
\end{enumerate}
The outline of the remainder of the paper is as follow. The preliminaries, problem definition and assumptions are provided in Section \ref{sec:problem}. In Section \ref{sec:exact-solution}, we discuss the theoretical aspects of our proposed fault estimation scheme. We present  the development and design of data-driven fault detection and isolation filters in Section \ref{sec:FDI}. Next, we propose a data-driven fault estimation filter for both the actuators and sensors as well as a tuning procedure is introduced and developed in Section \ref{sec:estimation}. Finally, we provide a number of illustrative simulation results in Section \ref{sec:simulation}.}

%%%%%%%%%%%%%%%%%%%%%%%%%%%%%%%%%%%%%%%%%%%%%%%%%%%%%%%%%%%%%%%%%%%%%%%%%%%%%%%%%%%%%%%%%%%%%%%%%%%%%%%%%%%%%%%%%%%%%%
\section{Preliminaries}\label{sec:problem}
Consider the following discrete-time linear system $\mathbf{S}$,
\begin{equation}\label{eq:system}
\mathbf{S}: \left\lbrace\begin{array}{l}
x(k+1)=Ax(k)+Bu(k)+w(k)\\y(k)=Cx(k)+v(k)
\end{array}\right.
\end{equation}
where $x \in \mathbb{R}^{n}$, $u \in \mathbb{R}^m$, and $y \in \mathbb{R}^l$. Moreover, $w (k) \in \mathbb{R}^n$ and $v(k) \in \mathbb{R}^l$ are white noise having zero mean and covariance matrices:
\begin{equation}
\mathbf{E}[\left[\begin{array}{c} w_i\\v_i\end{array}\right]\left[\begin{array}{cc} w_j^T&v_j^T\end{array}\right]]=\left[\begin{array}{cc} Q&S\\S^T&R\end{array}\right]\delta_{i,j}
\end{equation}
We model a given actuator or a sensor fault through additive terms that are injected in the system $\mathbf{S}$ as follows,
\begin{equation}\label{eq:faulty-system}
\mathbf{S}_f: \left\lbrace\begin{array}{l}
x(k+1)=Ax(k)+Bu(k)+Bf^a(k)+w(k)\\y(k)=Cx(k)+f^s(k)+v(k)
\end{array}\right.
\end{equation}
where $f^a(k) \in \mathbb{R}^m$ and $f^s(k)\in \mathbb{R}^l$ represent the actuator and sensor faults, respectively. These faults are commonly known as  \textit{additive faults}.
\begin{rem}
The actuator and sensor faults are traditionally modeled in various manners in the literature. For instance, either as additive faults or multiplicative faults. The proper choice depends on the actual characteristics of the fault. Typically, sensor bias, actuator bias and actuator loss of effectiveness are considered as additive faults. Multiplicative fault models are more suitable for representing changes in the system dynamic parameters such as gains and time constants \cite{patton2013issues}.
\end{rem}
{\textbf{Problem Statement}: The problem considered in this work deals with developing and designing fault detection, isolation and estimation schemes for \textit{both} sensor and actuator faults under the following assumptions.}

{ \textit{Assumption 1}: The system $\mathbf{S}$ is stable and observable.}

{ \textit{Assumption 2}: The system matrices and the system order are not known  \textit{a priori}.}

{ \textit{Assumption 3}: A sequence of \underline{healthy} measured system I/O data, namely $u(k)$ and $y(k)$, for $k=1, \ldots, T$, are available and the input $u(k)$ satisfies the persistently exciting (PE) condition \cite{ljung1998system}.}

{ \textit{Assumption 4}: The faults in the system $\mathbf{S}_f$ are \textit{detectable} and \textit{isolable}, as comprehensively discussed in \cite{ding2008model}.}

{The above assumptions are required in all the lemmas and theorems provided in the paper, however they are not explicitly stated in lemmas and theorems statements for sake of brevity. }

\textbf{Identification of the Markov Parameters}: We define the set $\{H_0,H_1,H_2, \ldots\}$, where $H_\beta=CA^{\beta}B$ is known as the \underline{Markov parameter}. If $u(k)$ is persistently exciting, then  several approaches are available in the literature  to directly identify the Markov parameters from the I/O data  $u(k)$ and $y(k)$ (\cite{dong2012identification,ljung1998system}). Specifically, we use \textit{correlation analysis} \cite{ljung1998system} for accomplishing the Markov parameters estimation task. The estimated Markov parameters are denoted by $\hat{H}_\beta$ in our subsequent derivations.

\textbf{Notation}: We will subsequently use an equivalent form of the system $\mathbf{S}$ as follows,
\begin{equation}\label{eq:system-subspace}
\mathbf{S}: \left\lbrace\begin{array}{l}
x(k-i+1)=Ax(k-i)+B\mathbf{I}_l^m\mathbf{U}_i(k-i)+w(k-i)\\ \mathbf{Y}(k-i)=\mathbf{C}x(k-i)+\mathbf{D}\mathbf{U}(k-i)\\ +\mathbf{E}\mathbf{W}(k-i)+\mathbf{V}(k-i)
\end{array}\right.
\end{equation}
where,
\begin{multline}\label{eq:cndnDef}
\mathbf{C}=\left(\begin{array}{c}C\\CA\\ \vdots \\ CA^{i}\end{array}\right);\mathbf{D}=\left(\begin{array}{ccccc} 0&0& \ldots &0& 0\\ H_0& 0& \ldots &0&0 \\ \vdots & \vdots & \vdots & \vdots&\vdots \\ H_{i-1}&H_{i-2}&\ldots & H_0&0\end{array}\right)\\ \mathcal{D}=\left[\begin{array}{cccc}
0 & H_1^T & \ldots & H_{i-1}^T \end{array}\right]^T \nonumber
\end{multline}
\begin{equation}
\mathbf{E}=\left(\begin{array}{ccccc} 0&0& \ldots & 0&0\\ C& 0& \ldots &0 &0\\ \vdots & \vdots & \vdots & \vdots&\vdots \\ CA^{i-1}&CA^{i-2}&\ldots &C& 0\end{array}\right)
\end{equation}

For any given signal $g(k)$, the following matrices are defined,
\begin{equation}
\mathbf{G}(k-i)=\left[\begin{array}{c}g(k-i) \\ g(k-i+1) \\ \vdots \\ g(k)
\end{array}   \right];\mathbf{G}_+(k-i)=\left[\begin{array}{c}g(k-i) \\ g(k-i+1) \\ \vdots \\ g(k+1)
\end{array}   \right]
\end{equation}

We extensively use the notation $\mathbf{I}_\alpha^\gamma$ which is defined as,
\begin{equation}
\mathbf{I}^\alpha_\gamma=\left[\begin{array}{cc} \mathbf{I}_{\alpha \times \alpha} & \mathbf{0}_{\alpha \times (i\gamma-\alpha)}
\end{array}\right]
\end{equation}
 Moreover, we also define,
\begin{multline}
\mathbf{D}_{+}=\left(\begin{array}{ccccc} H_0& 0& \ldots &0&0 \\ \vdots & \vdots & \vdots & \vdots&\vdots \\ H_{i}&H_{i-1}&\ldots &H_0& 0 \end{array}\right); \mathcal{D}_{+}=\left(\begin{array}{c} H_0 \\ \vdots  \\ H_{i} \end{array}\right)\\ \mathbf{C}_+=\left[\begin{array}{cccc}
(CA)^T & \ldots & (CA^{i+1})^T \end{array}\right]^T
\end{multline}
\begin{multline}\label{eq:u-hankel}
\mathbf{G}_{i,j}(k-i)=\\
\left(\begin{array}{cccc}g(k-i)&g(k-i+1)&\ldots&g(k-i+j)\\g(k-i+1)&g(k-i+2)&\ldots&g(k-i+j+1)\\ \vdots&\vdots&\vdots&\vdots \\g(k)&g(k+1)&\dots&g(k+j)\end{array}\right)
\end{multline}

{The matrices $\hat{\mathbf{D}}$, $\hat{\mathbf{D}}_+$, $\hat{\mathcal{D}}$ and $\hat{\mathcal{D}}_+$ are constructed similar to ${\mathbf{D}}$, ${\mathbf{D}}_+$, ${\mathcal{D}}$ and ${\mathcal{D}}_+$ where the actual Markov parameters $H_\beta$ are replaced by their estimates $\hat{H}_\beta$. }

{We define two sets $p$ and $q$ that contain a selection of integer numbers from 1 to $l$ and from 1 to $m$, respectively. The parameters $n_p$ and $n_q$ denote the number of elements in the sets $p$ and $q$, respectively. Both $p$ and $q$ can be empty sets denoted by $p=\{\emptyset\}$ and $q=\{\emptyset\}$. We denote by $kp$ (or $kq$), $k \in \mathbb{N}$, as a set that contains all the elements of $p$ (or $q$) multiplied by $k$. The notation $\sim q$ (or $\sim p$) denotes the set that contains the integers from 1 to $m$ (or 1 to $l$) that are not included in $q$ (or $p$). For example, for a given Markov parameter matrix $H_0 \in \mathbb{R}^{5\times 4}$, a typical $p$ can be taken as $p=\{2,4\}$. Moreover, $n_p=2$, $3p=\{6,12\}$ and $\sim p=\{1,3,5\}$.  If $p=\{\emptyset\}$, then $\sim p =\{1,2,3,4,5\}$. The matrix $\mathbf{O}_{p-}^{q-}$ is obtained by deleting the \underline{columns} $q,2q,\ldots,iq$ and \underline{rows} $p,2p,\ldots, ip$ of $\mathbf{O}$, respectively. The matrix $\mathbf{O}^{q+}$ and $\mathbf{O}_{p+}$ are defined as matrices that only contain the \underline{columns} $q,2q,\ldots,iq$ and the \underline{rows} $p,2p,\ldots, ip$ of $\mathbf{O}$, respectively.  The vector $\mathbf{P}^{p-}$ is obtained by deleting the \underline{rows} $p,2p,\ldots, ip$ of $\mathbf{P}$. Finally,  $\mathbf{P}^{p+}$ only contains the \underline{rows} $p,2p,\ldots, ip$ of $\mathbf{P}$. Similar notations are defined for $\mathbf{P}^{q+}$ and $\mathbf{P}^{q-}$. The signs $\dagger$, $\mathcal{N}$ and $\mathbb{E}\{.\}$  denote the Moore-Penrose pseudo inverse, null space and the expectation operator. The matrix $\mathbf{O}(\alpha:\beta,\gamma:\theta)$ denotes a matrix that is constructed from an original matrix $\mathbf{O}$ by only containing  the rows $\alpha$ to $\beta$ and the columns $\gamma$ to $\theta$ . If  $\alpha$ and $\beta$ (or $\gamma$ and $\theta$) are not specified, then it implies that we are dealing with all the rows (or columns) of $\mathbf{O}$.
\begin{rem}
The parameters $p$ and $q$ are defined in order to specify the set of I/O data that is to be fed to a fault isolation or estimation filter. For example, for a given Markov parameter matrix $H_0 \in \mathbb{R}^{5\times 4}$, one may desire to design a filter that operates with information from input channels $\{3,4\}$ and measurement channels $\{1,2,3\}$. Then, one should set $q=\{1,2\}$ and $p=\{4,5\}$. The above notation is critical for the task of fault isolation where one requires to construct a bank of filters each of which operates with a different set of inputs and outputs data. 
\end{rem}}

%%%%%%%%%%%%%%%%%%%%%%%%%%%%%%%%%%%%%%%%%%%%%%%%%%%%%%%%%%%%%%%
{
\section{Proposed Fault Estimation Scheme using Exact Markov Parameters and Observability Matrix}\label{sec:exact-solution}
In this section, we start by assuming availability of the exact Markov parameters and the extended observability matrix to introduce the basic concepts we utilize in this work. These assumptions will be relaxed and removed in the subsequent Sections \ref{sec:FDI}, \ref{sec:estimation} and \ref{sec:simulation}.\\}

{Let us consider a signal $\eta(k)$ that is governed by the following dynamics and stimulated by the information from the sensors $\sim p$ and actuators $\sim q$, that is
 \begin{equation} \label{eq:eta-def}
\eta(k+1)=A_r\eta(k)+B_r\mathbf{U}^{q-}(k-i)+L_r\mathbf{Y}^{p-}(k-i)
\end{equation}
 where $\eta(k) \in \mathbb{R}^{il'}$, $\mathbf{U}^{q-}(k-i) \in \mathbb{R}^{im'}$ and $\mathbf{Y}^{p-} \in \mathbb{R}^{il'}$, where $l'=(l-n_p)$ and $m'=(m-n_q)$. Our goal is to determine the \textit{unknown} matrices $A_r$, $B_r$ and $L_r$ such that for the healthy system $\mathbf{S}$ given by (\ref{eq:system-subspace}), we have
\begin{multline}\label{eq:estimation-error-condition}
\mathbb{E}(e(k))=\mathbb{E}(\eta(k)-\mathbf{T}x(k-i)) \rightarrow 0 \mbox{ as }k \rightarrow \infty
\end{multline}
where $\mathbf{T} \in \mathbb{R}^{l' \times n}$ denotes a full column rank matrix. The error dynamics associated with $e(k)$ is therefore given by,
\begin{eqnarray}\label{eq:eta-x}
e(k+1)&=&\eta(k+1)-\mathbf{T}x(k-i+1) \nonumber \\
&=& A_r e(k)+(A_r\mathbf{T}-\mathbf{T}A+L_r\mathbf{C}_{p-})x(k-i)\nonumber \\ &+&(B_r+L_r\mathbf{D}_{p-}^{q-}-\mathbf{T}B^{q-}\mathbf{I}_{l'}^{m'})\mathbf{U}^{q-}(k-i) \nonumber \\
&+&(L_r\mathbf{D}_{p-}^{q+}-\mathbf{T}B^{q+}\mathbf{I}_{l'}^{n_q})\mathbf{U}^{q+}(k-i) \nonumber \\
&+&L_r\mathbf{E}_{p-}^{q-}\mathbf{W}^{q-}(k-i)+L_r\mathbf{V}^{p-}(k-i)
\end{eqnarray}
which is obtained by substituting $\eta(k+1)$  from equation (\ref{eq:eta-def}) and $x(k-i+1)$ from equation (\ref{eq:system-subspace}). Condition (\ref{eq:estimation-error-condition}) is now satisfied if and only if (\textbf{a}) $A_r$ is a Hurwitz matrix, (\textbf{b}) $A_r\mathbf{T}-\mathbf{T}A+L_r\mathbf{C}_{p-}=0$, (\textbf{c}) $B_r+L_r\mathbf{D}_{p-}^{q-}-\mathbf{T}B^{q-}\mathbf{I}_{l'}^{m'}=0$, and (\textbf{d}) $L_r\mathbf{D}_{p-}^{q+}-\mathbf{T}B^{q+}\mathbf{I}_{l'}^{n_q} =0$. The above conditions actually correspond to the Luenberger observer equations.}

{The key concept that is introduced in this paper is that we specifically set,
\begin{equation}\label{eq:tci}
\mathbf{T}=\mathbf{C}_{p-}
\end{equation}
In other words, we select $\mathbf{T}$ to be equal to the extended observability matrix. Let us now define the matrix $\mathbf{M}_{p-}$ as follows,
\begin{equation}\label{eq:M-definition-org}
\mathbf{M}_{p-}=A_r+L_r
\end{equation}
Given (\ref{eq:tci}) and (\ref{eq:M-definition-org}) and in view of the fact that $\mathbf{C}_{p-}B^{q-}=\mathcal{D}_{+,p-}^{q-}$, $\mathbf{C}_{p-}B^{q+}=\mathcal{D}_{+,p-}^{q+}$, and $\mathbf{C}_{p-}A=\mathbf{C}_{+,p-}$, the conditions (\textbf{a}) to (\textbf{d}) can be rewritten as,
\begin{eqnarray}
{A}_r&& \mbox{ is Hurwitz}\label{eq:eq1org}\\
\mathbf{M}_{p-}\mathbf{C}_{p-}&=&\mathbf{C}_{+,p-} \label{eq:eq2org}\\
L_r{\mathbf{D}}_{p-}^{q+}-{\mathcal{D}}_{+,p-}^{q+}\mathbf{I}_{l'}^{n_q}&=&0 \label{eq:eq3org} \\
B_r+L_r{\mathbf{D}}_{p-}^{q-}-{\mathcal{D}}_{+}^{q-}\mathbf{I}_{l'}^{m'}&=&0 \label{eq:eq4org}
\end{eqnarray}
\begin{rem}
Recall that $\mathbf{C}_{p-}$ should be full column rank according to the assignment (\ref{eq:tci}). The matrix $\mathbf{C}_{p-}$ will be full column rank if $i$ is selected to be equal to or greater than the observability index of the pair $(C^{p-}, A)$, which is denoted by $\nu_p$. Given $i \geq \nu_p$, the matrix $\mathbf{M}_{p-}$ is given by $\mathbf{C}_{+,p-}(\mathbf{C}_{p-})^\dagger+\Theta(\mathbf{I}-\mathbf{C}_{p-}(\mathbf{C}_{p-})^\dagger)$,  where $\Theta$ is an arbitrary matrix introduced due to the Moore-Penrose pseudo inverse non-unique solution.
\end{rem}
\begin{defn}
The \textit{relative degree} of the subsystem inputs to the outputs $\sim p$ is defined as the smallest non-negative $\tau_p$ such that $H_{i,p-}=0$ for $i<\tau_p$ and $\mathcal{D}_{p-}$ is full column rank for $i \geq \tau_p$ \cite{wan2015data,kirtikar2009delay}.
\end{defn}
\begin{rem}
It is well-known that the system of equations (\ref{eq:M-definition-org}) and (\ref{eq:eq1org}) to (\ref{eq:eq4org}) has a solution if and only if i) $i \geq \tau_p$ and ii) the subsystem from the inputs $q$ to the outputs $\sim p$ is minimum phase \cite{UIO}. Particularly, the above equations always have a solution if $q=\{\emptyset\}$ since equation (\ref{eq:eq3org}) vanishes. Consequently, one can arbitrarily select a Hurwitz $A_r$ and then calculate $L_r$ and $B_r$ from equations (\ref{eq:M-definition-org}) and (\ref{eq:eq4org}).  The restriction that is imposed is actually on the subsystems and not the entire system. Therefore, the designer has a freedom to select a different $q$ if the minimum phase condition is not satisfied for the original selection.  
\end{rem}
Let us assume that it is desired to estimate the faults in the actuators $q$ by using the information from the sensors $\sim p$ and actuators $\sim q$. Then, the fault estimator filter is given by,
\begin{equation}\label{eq:actuator-est-org}
\left\lbrace\begin{array}{l}
\eta(k+1)={A}_r\eta(k)+{B}_r\mathbf{U}^{q-}(k-i)+{L}_r\mathbf{Y}^{p-}(k-i)\\ \hat{f}^a(k-i)=-\mathbf{I}_{m}^m\mathbf{D}^\dagger\left(\eta(k)-\right. \\ \left.\mathbf{Y}^{p-}(k-i)+{\mathbf{D}}_{p-}\mathbf{U}(k-i)\right)
\end{array}\right.
\end{equation}
where $\hat{f}^a(k)$ denotes an estimate of $f^a(k)$. The matrices $A_r$, $B_r$ and $L_r$ are obtained by solving the equations (\ref{eq:M-definition-org}) and (\ref{eq:eq1org}) to (\ref{eq:eq4org}).
\begin{thm}\label{thm:actuator-bias-org}
Assume that the subsystem from the inputs $q$ to the outputs $\sim p$ is minimum phase, $i \geq \max\{\nu_p,\tau_p\}$ and the sensors $\sim p$ and actuators $\sim q$ are healthy, then the filter dynamics governed by (\ref{eq:actuator-est-org}) is asymptotically unbiased.
\end{thm}
\begin{pf}
The proof is provided in the Appendix \ref{app:actuator-bias-org}. $\blacksquare$
\end{pf}
The above theorem guarantees that our proposed filter will generally have a lower order than the estimation filter that is proposed in \cite{wan2015data} which requires $i \geq \nu_p+\tau_p$ or $i \rightarrow \infty$ depending on the transmission zeros of the quadruple $(A,B^{q-},\mathbf{C}_{p-},\mathcal{D}_{p-}^{q-})$. \\}

{The case of the sensor fault estimation is slightly different. One can estimate the faults in the sensors $p$ if the sensors $\sim p$ and all the actuators are healthy. The fault estimator is now given by,
\begin{equation}\label{eq:sensor-est-org}
\left\lbrace\begin{array}{l}
\eta(k+1)={A}_r\eta(k)+{B}_r\mathbf{U}(k-i)+{L}_r^{p\downarrow}\mathbf{Y}(k-i)\\ \hat{f}^s(k-i)=\mathbf{I}_l^l\left(\eta(k)-\mathbf{Y}^{p-}(k-i)+{\mathbf{D}}_{p-}\mathbf{U}(k-i)\right)
\end{array}\right.
\end{equation}
where $\hat{f}^s(k)$ denotes an estimate of $f^s(k)$, $L_r^{p\downarrow}$ denotes a matrix where its columns $p,2p,\ldots,ip$ are zero. The matrices $A_r$, $B_r$ and $L_r^{p\downarrow}$ are obtained by solving the equations (\ref{eq:M-definition-org}) and (\ref{eq:eq1org}) to (\ref{eq:eq4org}) by setting $p=q=\{\emptyset\}$ and by replacing $L_r$ with $L_r^{p\downarrow}$. }

{The above proposed filter (\ref{eq:sensor-est-org}) is unbiased as established by the following theorem.
\begin{thm}\label{thm:sensor-bias-org}
Assume that $i \geq \nu_p$ and the sensors $\sim p$ and all the actuators are healthy, then the filter dynamics that is governed by (\ref{eq:sensor-est-org}) is asymptotically unbiased.
\end{thm}
\begin{pf}
The proof is provided in the Appendix \ref{app:sensor-bias-org}. $\blacksquare$
\end{pf}
Note that the subsystem from the inputs to the outputs $\sim p$ is \textit{not} required to be minimum phase for solving the sensor fault estimation problem. Moreover, it can be theoretically shown that the filter (\ref{eq:sensor-est-org}) is unbiased by using information from the faulty actuators provided it is modified by the estimations that are provided by the filter (\ref{eq:actuator-est-org}). However, this coupling will cause significant biases in the data-driven solution since the actuator fault estimation scheme is itself biased.} 

%%%%%%%%%%%%%%%%%%%%%%%%%%%%%%%%%%%%%%%%%%%%%%%%%%%%%%%%%%%%%%%%%%%%%%%%%%%%%%%%%%%%%%%%%%%%%%%%%%%%%%%%%%%%%
{
\section{Data-driven Fault Detection and Isolation (FDI) Scheme}\label{sec:FDI}
In this section, our proposed fault detection and isolation (FDI) filters are now directly constructed from the healthy system I/O data. First, we propose a data-driven estimation of the matrix $\mathbf{M}_{p-}$ and then present the design procedure of the FDI filters. 
\begin{rem}
Theorems \ref{thm:actuator-bias-org} and \ref{thm:sensor-bias-org} provide the guidelines for selection of $i$. The parameter $i$ is bounded by $n$ which is not known \textit{a priori}. The condition $i \geq \tau_p$ can be easily satisfied by checking the rank of $\mathcal{D}_{p-}$. However, the parameter $\nu_p$ is not known. Therefore, $i$ should be selected sufficiently large that ensures $i \geq \max\{\nu_p,\tau_p\}$.
\end{rem}
\subsection{Data-driven Estimation of the Filter Parameters}
In order to solve equations (\ref{eq:M-definition-org}) and (\ref{eq:eq1org}) to (\ref{eq:eq4org}), one requires the Markov parameters and the extended observability matrix. The extended observability matrix is required to obtain $\mathbf{M}_{p-}$. However, in our subsequent data-driven derivations we will show that an estimate of the matrix $\mathbf{M}_{p-}$ can be directly obtained from the system I/O data \textit{without} applying the reduction step. Consequently, the matrix $\mathbf{C}_{p-}$ or its equivalent forms are not actually required. \\}

{The objective of the equation (\ref{eq:eq2org}) is in fact to enforce,
\begin{equation}\label{eq:enforce-x-to-zero}
(A_r\mathbf{C}_{p-}-\mathbf{C}_{p-}A+L_r\mathbf{C}_{p-})x(k-i) \equiv 0
\end{equation}
On the other hand, from the measurement equation (\ref{eq:system-subspace}) it follows that,
\begin{multline}\label{eq:cx-versus-y-du}
\mathbf{C}_{p-}x(k-i)=\mathbf{Y}^{p-}(k-i)-\mathbf{D}_{p-}\mathbf{U}(k-i)\\ -\mathbf{E}_{p-}\mathbf{W}(k-i)-\mathbf{V}^{p-}(k-i)
\end{multline}
By substituting equation (\ref{eq:cx-versus-y-du}) into equation (\ref{eq:enforce-x-to-zero}) one obtains,
\begin{eqnarray} \label{eq:m-equation}
&&\mathbf{M}_{p-}(\mathbf{Y}^{p-}(k-i)-\mathbf{D}_{p-}\mathbf{U}(k-i)-\mathbf{E}_{p-}\mathbf{W}(k-i)\nonumber \\ &-&\mathbf{V}^{p-}(k-i))
-(\mathbf{Y}^{p-}(k-i+1)-\mathbf{D}_{+,p-}\mathbf{U}_{+}(k-i) \nonumber \\ &-&\mathbf{E}_{+,p-}\mathbf{W}_{+}(k-i)-\mathbf{V}^{p-}(k-i+1)) \nonumber \\
&=&0
\end{eqnarray}
where,
\begin{equation}\label{eq:m-ar-lr}
\mathbf{M}_{p-}  \delequal A_r+L_r
\end{equation}
Iterating the equation (\ref{eq:m-equation})  from the time steps $k-i$ to $k-i+j$, where $j \gg i$, yields,
\begin{multline}\label{eq:m-equation-2}
\mathbf{M}_{p-}(\Gamma_0^{p-}-\mathbf{E}_{p-}\mathbf{W}_{i,j}(k-i)-\mathbf{V}^{p-}_{i,j}(k-i))\\ -(\Gamma_1^{p-}-\mathbf{E}_{+,p-}\mathbf{W}^{p-}_{(i+1),j}(k-i)-\mathbf{V}^{p-}_{(i+1),j}(k-i+1))=0
\end{multline}
where $\Gamma_0^{p-}=\mathbf{Y}^{p-}_{i,j}(k-i)-\mathbf{D}_{p-}\mathbf{U}_{i,j}(k-i)$ and $\Gamma_1^{p-}=\mathbf{Y}_{(i+1),j}^{p-}(k-i+1)-\mathbf{D}_{+,p-}\mathbf{U}_{(i+1),j}(k-i)$.  Equation (\ref{eq:m-equation-2}) forms the basis for our proposed data-driven solution for estimating $\mathbf{M}_{p-}$. The matrix $\mathbf{M}_{p-}$ minimizes the following cost function,
\begin{equation*}
\|\Gamma_1^{p-}-\mathbf{M}_{p-}\Gamma_0^{p-}\|_2
\end{equation*}}
{We do not have access to the actual values of the Markov parameters. Instead, we construct the matrices $\hat{\Gamma}_0^{p-}$ and $\hat{\Gamma}_1^{p-}$ by using the estimated Markov parameters and the system I/O data (healthy data) as follows,
\begin{equation}\label{eq:gamma0}
\hat{\Gamma}_0^{p-}=\mathbf{Y}_{i,j}^{p-}(k-i)-\hat{\mathbf{D}}_{p-}\mathbf{U}_{i,j}(k-i)
\end{equation}
\begin{equation}\label{eq:gamma1}
\hat{\Gamma}_1^{p-}=\mathbf{Y}_{(i+1),j}^{p-}(k-i+1)-\hat{\mathbf{D}}_{+,p-}\mathbf{U}_{(i+1),j}(k-i)
\end{equation}
where $\hat{\mathbf{D}}_{p-}$ and $\hat{\mathbf{D}}_{+,p-}$ are constructed similar to ${\mathbf{D}}_{p-}$ and ${\mathbf{D}}_{+,p-}$  but instead the estimated Markov parameters are utilized. Therefore, an estimate of $\mathbf{M}_{p-}$ is given by $\hat{\Gamma}_1^{p-} (\hat{\Gamma}_0^{p-})^\dagger+\Theta(\mathbf{I}-\hat{\Gamma}_0^{p-} (\hat{\Gamma}_0^{p-})^\dagger)$, where $\Theta$ is an arbitrary matrix. However, the solution will be unique as $j \rightarrow \infty$ as stated in the following lemma.
\begin{lem}\label{lem:m-consitent}
If $j \rightarrow \infty$, then the matrix $\hat{\Gamma}_0^{p-}$ is full row rank and,
\begin{equation}\label{eq:m-solution}
\hat{\mathbf{M}}_{p-}= \hat{\Gamma}_1^{p-} (\hat{\Gamma}_0^{p-})^\dagger 
\end{equation}
\end{lem}
\begin{pf}
The proof is provided in the Appendix \ref{app:gamma-rank}. $\blacksquare$
\end{pf}
The matrix $\hat{\mathbf{M}}_{p-}$ has a particular structure as shown in the following lemma.
\begin{lem}\label{lem:mhat-structure}
The matrix $\hat{\mathbf{M}}_{p-}$ has the following structure,
\begin{equation}\label{eq:mhat-struct}
\hat{\mathbf{M}}_{p-} = \left[\begin{array}{cc} 0_{(i-1)l' \times l'}& \mathbf{I}_{(i-1)l'\times(i-1)l' }\\ \mathbf{K}_1 & \mathbf{K}_2\end{array}\right]
\end{equation}
where $\mathbf{K}_1 \in \mathbb{R}^{l' \times l'}$ and $\mathbf{K}_2 \in \mathbb{R}^{l'\times(i-1)l'}$ are nonzero matrices, where $l'=l-n_p$.
\end{lem}
\begin{pf}
One can partition $\hat{\Gamma}_0^{p-}$ and $\hat{\Gamma}_1^{p-}$ as follows,
\begin{equation}
\hat{\Gamma}_0^{p-}=\left[\begin{array}{c}\hat{\Gamma}_{01} \\ \hat{\Gamma}_{02}\end{array}\right]; \mbox{ } \hat{\Gamma}_1^{p-}=\left[\begin{array}{c}\hat{\Gamma}_{11} \\ \hat{\Gamma}_{12}\end{array}\right]
\end{equation}
where $\hat{\Gamma}_{01} \in \mathbb{R}^{l' \times il'}$, $\hat{\Gamma}_{02} \in \mathbb{R}^{(i-1)l' \times il'}$, $\hat{\Gamma}_{11} \in \mathbb{R}^{l' \times il'}$ and $\hat{\Gamma}_{12} \in \mathbb{R}^{(i-1)l' \times il'}$. It follows readily from definitions of  $\hat{\Gamma}_0^{p-}$ and $\hat{\Gamma}_1^{p-}$ (equations (\ref{eq:gamma0}) and (\ref{eq:gamma1})) that $\hat{\Gamma}_{11}=\hat{\Gamma}_{02}$. Since $\hat{\Gamma}_0^{p-}$ is full row rank, therefore we have
\begin{equation}
\hat{\Gamma}_{11}=\left[\begin{array}{cc}0 & \mathbf{I}\end{array}\right]\left[\begin{array}{c}\hat{\Gamma}_{01} \\ \hat{\Gamma}_{02}\end{array}\right]
\end{equation}
Moreover, $\hat{\Gamma}_{12}$ can be written as a linear combination of the rows of $\hat{\Gamma}_{01}$ and $\hat{\Gamma}_{02}$ as follows,
\begin{equation}
\hat{\Gamma}_{12}=\left[\begin{array}{cc}\mathbf{K}_1 & \mathbf{K}_2\end{array}\right]\left[\begin{array}{c}\hat{\Gamma}_{01} \\ \hat{\Gamma}_{02}\end{array}\right]
\end{equation}
Consequently, we obtain,
\begin{equation}
\hat{\Gamma}_{1}^{p-}=\left[\begin{array}{cc}0 & \mathbf{I} \\ \mathbf{K}_1 & \mathbf{K}_2\end{array}\right]\hat{\Gamma}_{0}^{p-} 
\end{equation}
which reveals the general structure of $\hat{\mathbf{M}}_{p-}$ as given by equation (\ref{eq:mhat-struct}). $\blacksquare$
\end{pf}
\begin{lem}\label{lem:mhat-hurwitz}
The matrix $\hat{\mathbf{M}}_{p-}$ is Hurwitz.
\end{lem}
\begin{pf}
The proof is provided in the Appendix \ref{app:m-hurt}. $\blacksquare$
\end{pf}
The above analysis shows that the estimation filter (\ref{eq:eta-def}), which satisfies the condition (\ref{eq:estimation-error-condition}), can be directly synthesized from the system I/O data \textit{without} requiring any reduction step. The data-driven counterparts of equations (\ref{eq:M-definition-org}) and (\ref{eq:eq1org}) to (\ref{eq:eq4org}) is now given by,
\begin{eqnarray}
\hat{A}_r&& \mbox{ is Hurwitz}\label{eq:eq1}\\
\hat{A}_r+\hat{L}_{r}&=&\hat{\mathbf{M}}_{p-} \label{eq:eq2}\\
\hat{L}_r\hat{\mathbf{D}}_{p-}^{q+}-\hat{\mathcal{D}}_{+,p-}^{q+}\mathbf{I}_{l'}^{n_q}&=&0 \label{eq:eq3} \\
\hat{B}_r+\hat{L}_r\hat{\mathbf{D}}_{p-}^{q-}-\hat{\mathcal{D}}_{+}^{q-}\mathbf{I}_{l'}^{m'}&=&0 \label{eq:eq4}
\end{eqnarray}
\begin{thm}\label{thm:sol-exist}
Equations (\ref{eq:eq1}) to (\ref{eq:eq4}) have a solution if and only if $\hat{M}^{p-}-\hat{\mathcal{D}}_{+,p-}^{q+}\mathbf{I}_{l'}^{n_q}(\hat{\mathbf{D}}_{p-}^{q+})^\dagger-\Theta_1(\mathbf{I}-\hat{\mathbf{D}}_{p-}^{q+}(\hat{\mathbf{D}}_{p-}^{q+})^\dagger)$ is Hurwitz for an arbitrary matrix $\Theta_1 \in \mathbb{R}^{il' \times il'}$.
\end{thm}
\begin{pf}
Solving equation (\ref{eq:eq3}) for $\hat{L}_r$ yields,
\[\hat{L}_r=\hat{\mathcal{D}}_{+,p-}^{q+}\mathbf{I}_{l'}^{n_q}(\hat{\mathbf{D}}_{p-}^{q+})^\dagger+\Theta_1(\mathbf{I}-\hat{\mathbf{D}}_{p-}^{q+}(\hat{\mathbf{D}}_{p-}^{q+})^\dagger)\]
Substituting the above expression in equation (\ref{eq:eq2}) and comparing it with equation (\ref{eq:eq1}) concludes the result.  $\blacksquare$
\end{pf}
The arbitrary matrix $\Theta_1$ in $\hat{L}_r$ should be selected such that equation (\ref{eq:eq1}) is satisfied. Note that equation (\ref{eq:eq3}) and the free parameter $\Theta_1$ vanish when $q=\{\emptyset\}$. Therefore, one can arbitrarily select a Hurwitz matrix $\hat{A}_r$ and then obtain $\hat{L}_r$ and $\hat{B}_r$ from equations (\ref{eq:eq3}) and (\ref{eq:eq4}).}
% Theoretically, for exact values of the Markov parameters and the matrix $\mathbf{M}_{p-}$, a solution exists to equations (\ref{eq:eq1}) to (\ref{eq:eq4}) if and only if the subsystem from the inputs $q$ to $y^{p-}(k)$ is minimum phase \cite{UIO}. It implies that one can not obtain a Hurwitz $\hat{A}_r$ by any choice for $\Theta_1$. However, this property cannot be explicitly investigated since the system matrices are not known \textit{a priori}. In the data-driven context, the estimated Markov parameters and $\hat{\mathbf{M}}_{p-}$ are biased. Therefore, one may actually obtain a solution by adjusting $\Theta_1$ while the subsystem from the inputs $q$ to $y^{p-}(k)$ is in fact non-minimum phase. The above drawback appears if one is interested in designing an actuator fault isolation or fault estimation filters. If $q=\{\emptyset\}$, it implies that the subsystem from none of the inputs to $y^{p-}(k)$ should be minimum phase, which is always true. Mathematically speaking, equation (\ref{eq:eq3}) vanishes when $q=\{\emptyset\}$. Therefore, one can arbitrarily select a Hurwitz matrix $\hat{A}_r$ and then obtain $\hat{L}_r$ and $\hat{B}_r$ from equations (\ref{eq:eq3}) and (\ref{eq:eq4}).}   
%%%%%%%%%%%%%%%%%%%%%%%%%%%%%%%%%%%%%%%%%%%%%%%%%%%%%%%%%%
{
\subsection{Fault Detection and Isolation Filters}
Our proposed residual generator fault detection and isolation filter has the general structure that is as governed by,
\begin{equation}\label{eq:general-form}
\left\lbrace\begin{array}{l}
\eta(k+1)=\hat{A}_r\eta(k)+\hat{B}_r\mathbf{U}^{q-}(k-i)+\hat{L}_{r}\mathbf{Y}^{p-}(k-i)\\ \hat{r}(k)=\mathbf{I}_f(\eta(k)-\mathbf{Y}^{p-}_i(k-i)+\hat{\mathbf{D}}_{p-}\mathbf{U}(k-i))
\end{array}\right.
\end{equation}
where $\hat{r}(k)$ denotes the residual signal, $\mathbf{I}_f=\mathbf{I}_m^m$ if $q \neq \{\emptyset\}$ and $\mathbf{I}_f=\mathbf{I}$, otherwise. The filter parameters are obtained from equations (\ref{eq:eq1}) to (\ref{eq:eq4}).The above general structure can be configured for both single or concurrent fault detection, isolation or estimation tasks by invoking different settings for $p$ and $q$. Specifically, if both $p$ and $q$ are set to be empty sets, then the filter (\ref{eq:general-form}) will be in fact a fault detection filter.\\ }

{Fault isolation is typically performed via \textit{structured residuals}. In other words, a bank of residual observers is constructed where each filter in the bank is \underline{insensitive} to a particular fault but sensitive to all the other faults. Therefore, in case of occurrence of a fault, all the filters generate non-zero residuals that exceed their thresholds except for the one filter that can be used for determining the isolated fault. This can be achieved by invoking different settings for $p$ and $q$ for each filter in the bank.\\ }

{For example, if a single actuator fault isolation scheme for a system with three inputs and four measurements is desired, then a bank that consists of three filters should be constructed. A possible configuration setting for the filters 1,2 and 3 in the bank is $q=\{1,2\}$, $q=\{1,3\}$ and $q=\{2,3\}$, respectively, and $p=\{\emptyset\}$ for all. Alternatively, one may try the setting $q=\{1\}$, $p=\{3,4\}$, $q=\{2\}$, $p=\{1,2,4\}$, and $q=\{1,3\}$, $p=\{1,3\}$ for the filters 1, 2 and 3, respectively. A particular configuration selection depends on the context of the problem and the requirements. Despite the above flexibility, our proposed scheme has a limitation that it cannot handle simultaneous concurrent actuator \underline{and} sensor faults. This situation differs from the concurrent actuators \underline{or} concurrent sensors faults which is well managed within our proposed framework.}

{The residuals that are generated by the filter (\ref{eq:general-form}) has an important property that is characterized in the next lemma.
\begin{lem}\label{lem:trivial-zero}
Given $q=\{\emptyset\}$ and $\hat{A}_r$ selected to be a diagonal Hurwitz matrix, then the first $(i-1)l'$ rows of $\hat{r}(k)$ converge to zero as $k \rightarrow \infty$ independent of the presence of the faults.
\end{lem}
\begin{pf}
The proof is provided in the Appendix \ref{app:trivialZero}. $\blacksquare$
\end{pf}
Based on the above lemma, the first $(i-1)l'$ rows of $\hat{r}(k)$ do not contain any useful information that would allow a model reduction. The general structure of the residual generator filter for the actuator or sensor fault detection or the sensor fault isolation ($q=\{\emptyset\}$) is then given by,
\begin{equation}\label{eq:general-form-reduced}
\left\lbrace\begin{array}{l}
\eta_r(k+1)=\hat{\bar{A}}_r\eta_r(k)+\hat{\bar{B}}_r\mathbf{U}(k-i)+\hat{\bar{L}}_r\mathbf{Y}^{p-}(k-i)\\ \hat{\bar{r}}(k)=\eta_r(k)-y^{p-}(k)+\hat{\bar{\mathbf{D}}}_{p-}\mathbf{U}(k-i)
\end{array}\right.
\end{equation}
where $\eta_r(k) \in \mathbb{R}^{l'}$, $\hat{\bar{A}}_r={A}_r((i-1)l'+1:il,il'+1:il'$, $\hat{\bar{B}}_r=\hat{B}_r((i-1)l'+1:il',:)$, $\hat{\bar{L}}_r=\hat{L}_r((i-1)l'+1:il',:)$ and $\hat{\bar{\mathbf{D}}}_{i,p-}=\hat{\mathbf{D}}_{i,p-}((i-1)l'+1:il',:)$.\\}

{The same model reduction procedure cannot be applied to the actuator fault isolation filter (i.e. when $q \neq \{\emptyset\}$) since $\hat{A}_r$ that is obtained from equations (\ref{eq:eq1}) to (\ref{eq:eq4}) is not necessarily diagonal. }
% Ideally, the residuals of filter (\ref{eq:general-form}) (or reduced form (\ref{eq:general-form-reduced})) should converge to zero for a healthy system. However, the estimation error of Markov parameters and the Matrix $\hat{\mathbf{M}}$ cause errors that can be quite large. We address this issue by further tuning of the proposed residual generator filter. 
%%%%%%%%%%%%%%%%%%%%%%%%%%%%%%%%%%%%%%%%%%%%%%%%%%%%%%%%%%%%%%%%%%%%
{
\subsection{Residual Dynamics in Presence of a Fault}
Let us now investigate the corresponding residual dynamics in presence of faults. If $f^a(k)$ and/or $f^s(k)$ are nonzero, then the residual dynamics is given by,
\begin{equation}\label{eq:general-form-modif-fault-inject}
\left\lbrace\begin{array}{l}
\eta(k+1)=\hat{A}_r\eta(k)+\hat{B}_r\left(\mathbf{U}^{q-}(k-i)+\mathbf{F}^{a,q-}(k-i)\right)\\ +\hat{L}_{r}\left(\mathbf{Y}^{p-}(k-i)+\mathbf{F}^{s,p-}(k-i)\right)\\ \hat{r}(k)=\mathbf{I}_f \left(\eta(k)-\mathbf{Y}^{p-}_i(k-i)-\mathbf{F}^{s,p-}(k-i)\right. \\ \left.+\hat{\mathbf{D}}_{p-}(\mathbf{U}(k-i)+\mathbf{F}^{a}(k-i))\right)
\end{array}\right.
\end{equation}
where $\mathbf{F}^{a}(k-i)$ and $\mathbf{F}^{s}(k-i)$ are construed similar to $\mathbf{G}(k-i)$ using the actuator fault and the sensor fault signals, respectively. The vectors $\mathbf{F}^{a,q-}(k-i)$ and $\mathbf{F}^{s,p-}(k-i)$ are then obtained by deleting the rows $q,\ldots,iq$ and $p,\ldots,ip$ of $\mathbf{F}^{a}(k-i)$ and $\mathbf{F}^{s}(k-i)$, respectively. The filter dynamics (\ref{eq:general-form-modif-fault-inject}) shows that the residual $\hat{r}(k)$ is clearly affected by the faults except those in the sensors $p$ or the actuators $q$. \\}

{Conventionally, the following decision logic is utilized for performing the fault detection task, namely
\begin{equation}\label{eq:det-logic2}
\left\lbrace \begin{array}{l} \mbox{If } r_{min} \leq \mathbb{E}\{\|\hat{{r}}(k)\|\} \leq r_{max} \\  \Rightarrow \mbox{ System is healthy} \\  \mbox{If }\mathbb{E}\{\|\hat{{r}}(k)\|\}< r_{min} \mbox{ or } \mathbb{E}\{\|\hat{{r}}(k)\|\}> r_{max} \\ \Rightarrow  \mbox{ System is faulty}\end{array}\right.
\end{equation}
where $r_{min}$ and $r_{max}$ denote the lower and the upper bound thresholds, respectively. The thresholds are selected through conducting comprehensive Monte Carlo simulation runs so that the missed alarms and false alarms are minimized.\\}

{ A similar structure can be utilized for selecting the fault isolation decision logic. Specifically, a fault is detected and isolated in the actuator $q_0$ if,
\begin{equation}
\left\lbrace \begin{array}{ll} \mathbb{E}\{\|\hat{r}^a(k)\|\} < r_{min} \mbox{ or }\mathbb{E}\{\|\hat{r}^a(k)\|\} > r_{max} &\mbox{ ; } \\ a \neq q_0, a=1,\ldots, m, \mbox{ and } \\ r_{min} \leq \mathbb{E}\{\|\hat{r}^a(k)\|\} \leq r_{max} &\mbox{ ; } a=q_0.\end{array}\right.
\end{equation}
where $\hat{r}^a(k)$ denotes the residual that is obtained by setting $q=\{a\}$.\\}

{This completes our proposed solution to the problem of data-driven fault detection and isolation. In the next section, we consider the problem of data-driven fault estimation.}

%%%%%%%%%%%%%%%%%%%%%%%%%%%%%%%%%%%%%%%%%%%%%%%%%%%%%%%%%%%%%%%%%%%%%%%%%%%%%%%%%%%%%%%%%%%%%%%%%%%%%%%
{
\section{The Proposed Fault Estimation Scheme}\label{sec:estimation}
In many practical control problems, it is crucial to estimate the faults once they are detected and isolated. In this section, we provide a data-driven based methodology for design of fault estimation filters. Our proposed fault estimation scheme can be integrated with the FDI scheme. In other words, the FDI scheme introduced in the previous section can be utilized to distinguish between the healthy actuators and sensors from those where their data are used for fault estimation of faulty actuators and sensors. \\}

{In this section, we first propose fault estimation filters. It turns out that these filters are biased due to presence of estimation errors in the Markov parameters and the matrix $\hat{\mathbf{M}}_{p-}$. We then derive the dynamics corresponding to the fault estimation errors and show that it can be directly identified from the healthy system I/O data. Finally, we propose our so-called \textit{tuned fault estimation filters} that are obtained for a reliable and actuator fault estimation by integrating the proposed estimation filters with the identified estimation error dynamics.}
%%%%%%%%%%%%%%%
{\subsection{Sensor Fault Estimation Filters} \label{subsec:sen-est}
The following filter is now proposed to estimate the faults in the sensors $p$ by using data from the sensors $\sim p$ and all the actuators,
\begin{equation}\label{eq:sensor-est}
\left\lbrace\begin{array}{l}
\eta(k+1)=\hat{A}_r\eta(k)+\hat{B}_r\mathbf{U}(k-i)+\hat{L}_r^{p\downarrow}\mathbf{Y}(k-i)\\ \doublehat{f}^{s}(k-i)=\mathbf{I}_l^l \left(\eta(k)-\mathbf{Y}(k-i)+\hat{\mathbf{D}}\mathbf{U}(k-i)\right)
\end{array}\right.
\end{equation}
where $\doublehat{f}^{s}(k-i)$ denotes an estimate of $f^s(k-i)$ and the filter parameters are obtained from the equations (\ref{eq:eq1}) to (\ref{eq:eq4}) by setting $p=q=\{\emptyset\}$ and by replacing $\hat{L}_{r}$ with $\hat{L}_r^{p\downarrow}$.\\} 

{Clearly, the filter (\ref{eq:sensor-est}) is biased due to presence of estimation errors in the Markov parameters and the matrix $\hat{\mathbf{M}}$. The matrix $\hat{\mathbf{M}}$ is defined to be the same as $\hat{\mathbf{M}}_{p-}$ when $p=\{\emptyset\}$. Let us define the estimation error as $\Delta f^{s}(k)=f^{s}(k-i)-\doublehat{f}^{s}(k-i)$. Therefore, we have,
\begin{multline}
\Delta f^{s}(k-i)=\mathbf{I}_l^l\left(\xi(k)+\Delta \mathbf{D}\mathbf{U}(k-i)\right.\\ \left.+\mathbf{E}\mathbf{W}(k-i)+\mathbf{V}(k-i)\right)
\end{multline}
where $\xi(k)=\mathbf{C}x(k-i)-\eta(k)$ and $\Delta \mathbf{D}=\mathbf{D}-\hat{\mathbf{D}}$. The dynamics of $\xi(k)$ is now governed by,
\begin{eqnarray}\label{eq:zeta-x}
\xi(k+1)&=& \hat{A}_r \xi(k)+(\hat{A}_r-{\mathbf{M}}+\hat{L}_r^{p\downarrow})\mathbf{C}x(k-i)\nonumber \\ &+&(\hat{B}_r+\hat{L}_r^{p\downarrow}\mathbf{D}-\mathbf{C}B\mathbf{I}_{l}^{m})\mathbf{U}(k-i) \nonumber \\
&+&\hat{L}_r^{p\downarrow}\mathbf{E}\mathbf{W}(k-i)+\hat{L}_r^{p\downarrow}\mathbf{V}(k-i)
\end{eqnarray}
The matrix ${\mathbf{M}}$ is equal to ${\mathbf{M}}_{p-}$ when $p=\{\emptyset\}$. We substitute $\mathbf{C}x(k-i)$ by $\mathbf{Y}
(k-i)-\mathbf{D}\mathbf{U}(k-i)-\mathbf{F}^s(k-i)-\mathbf{E}\mathbf{W}(k-i)-\mathbf{V}(k-i)$ in the above equation. Rearranging of the right hand side of the above equation after substitution yields the governing dynamics of the fault estimation error as follows,
\begin{equation}\label{eq:residual-error-dyn}
\left\lbrace\begin{array}{l}
\xi(k+1)=\hat{A}_r\xi(k)+{\Delta}_u \mathbf{U}(k-i)+{\Delta}_y\mathbf{Y}(k-i)-\\{\Delta}_y\mathbf{F}^s(k-i)+\mathcal{M}_1 \mathbf{W}(k-i)+\mathcal{M}_2\mathbf{V}(k-i)\\ \Delta f^{s}(k-i)=\mathbf{I}_l^l\left(\xi(k)+\Delta \mathbf{D}\mathbf{U}(k-i)\right.\\ \left.+\mathbf{E}\mathbf{W}(k-i)+\mathbf{V}(k-i)\right)\end{array}\right.
\end{equation}
where $\delta A=\hat{A}_r-\mathbf{M}+\hat{L}_r^{p\downarrow}$, $\delta B=\hat{B}_r+\hat{L}_r^{p\downarrow}\mathbf{D}-\mathbf{C}B\mathbf{I}_{l}^{m}$, $\Delta_u=
\delta B-\delta A \mathbf{D}$, $\Delta_y=\delta A$, $\mathcal{M}_1=
\hat{L}_r^{p\downarrow}\mathbf{E}-\delta A\mathbf{E}$ and $\mathcal{M}_2=\hat{L}_r^{p\downarrow}\delta A$.\\}

{Equation (\ref{eq:residual-error-dyn}) clearly shows that the fault estimates are biased. All the parameters in the filter (\ref{eq:residual-error-dyn}) are unknown since their computation requires the exact Markov parameters and the matrix $\mathbf{M}$. However, we will show that one can actually obtain an estimate of $\Delta_u$, $\Delta_y$ and $\Delta\mathbf{D}$ by using the healthy I/O data.}

{Towards this end, we split the off-line available healthy I/O data into two segments. The first segment is utilized to estimate the system Markov parameters and the matrix $\hat{\mathbf{M}}$. Once the filter (\ref{eq:sensor-est}) is constructed, it is stimulated by the second segment of the I/O data to obtain $\doublehat{f}^{s}(k-i)$ using an arbitrary initial condition for $\eta (0)$ in (\ref{eq:sensor-est}). Theoretically, $\doublehat{f}^{s}(k-i)$ should be zero corresponding to the second segment of the healthy data, however, it will be biased due to presence of the estimation errors in the Markov parameters and the matrix $\hat{\mathbf{M}}$. The bias is governed and is given by $\Delta f^{s}(k)={f}^{s}(k-i)-\doublehat{f}^{s}(k-i) \equiv -\doublehat{f}^{s}(k-i)$, and according to equation (\ref{eq:residual-error-dyn}) is governed by,
\begin{equation}\label{eq:residual-error-dyn2}
\left\lbrace\begin{array}{l}
\xi(k+1)=\hat{A}_r\xi(k)+{\Delta}_u \mathbf{U}(k-i)+{\Delta}_y\mathbf{Y}(k-i)-\\+\mathcal{M}_1 \mathbf{W}(k-i)+\mathcal{M}_2\mathbf{V}(k-i)\\ \Delta f^{s}(k)=\mathbf{I}_l^l\left(\xi(k)+\Delta \mathbf{D}\mathbf{U}(k-i)\right.\\ \left.+\mathbf{E}\mathbf{W}(k-i)+\mathbf{V}(k-i)\right)\end{array}\right.
\end{equation}
One may consider the filter (\ref{eq:residual-error-dyn2}) as a stochastic LTI system that is described by the quadruple $\left(\mathcal{A},\mathcal{B},\mathcal{C},\mathcal{G}\right) \equiv \left(\hat{A}_r,\left[\begin{array}{cc}\Delta_u^T & \Delta_y^T\end{array}\right]^T,\mathbf{I}_l^l,\left[\begin{array}{cc}\Delta \mathbf{D}^T & 0\end{array}\right]^T\right)$. The process and measurement noise are given by $\mathcal{M}_1 \mathbf{W}(k-i)+\mathcal{M}_2\mathbf{V}(k-i)$ and $\mathbf{E}\mathbf{W}(k-i)+\mathbf{V}(k-i)$, respectively. The matrices $\mathcal{A}$ and $\mathcal{C}$ and the order of this system are already known. Given that $\Delta f^{s}(k)=-\doublehat{f}^{s}(k-i)$ corresponding to the second segment of the healthy data, and $\mathcal{A}$ and $\mathcal{C}$, one can \textit{estimate} $\mathcal{B}$ and $\mathcal{G}=\left[\begin{array}{cc}\mathcal{G}_1 &0\end{array}\right]$ by invoking an optimization problem that is described in detail below.\\ }

{Let us define the matrices $\mathcal{E}_{1,j}(k)$ and $\mathbf{\mathcal{Z}}_{(\lambda-1),j}(k)$ similar to $\mathbf{G}_{i,j}(k)$ by replacing the signal $g(k)$ with the signals $-\doublehat{f}^{s}(k-i)$ and $\mathbf{Z}(k)$, respectively, where $\mathbf{Z}(k)=\left[\begin{array}{c}\mathbf{U}(k)\\\mathbf{Y}(k)\end{array}\right]$, and $j$ is selected as large as the available data of the second segment allows. Therefore, for the system (\ref{eq:residual-error-dyn2}), we have,
\begin{equation}\label{eq:ep-tau}
\mathcal{E}_{1,j}(k)=\mathcal{C}\mathcal{A}^\lambda \xi (k-\lambda)+\mathcal{T}_{1,\lambda}\mathbf{\mathcal{Z}}_{(\lambda-1),j}(k-\lambda)+ST
\end{equation}
where the term $ST$ denotes the stochastic terms which have zero mean and are neglected here for sake of brevity, and  $\mathcal{T}_{1,\lambda}$ is defined as,
\begin{multline} \label{eq:dsi-form}
\mathcal{T}_{1,\lambda}= \left(\begin{array}{cccccc} \mathcal{H}_{\lambda-1}&\mathcal{H}_{\lambda-2}& \ldots & \mathcal{H}_{0}& \mathcal{G} \end{array} \right)
\end{multline}
where $\mathcal{H}_\beta=\mathcal{C}\mathcal{A}^\beta\mathcal{B}$. The definition (\ref{eq:dsi-form}) shows dependence of $\mathcal{T}_{1,\lambda}$ on the matrices $\mathcal{B}$ and $\mathcal{G}$. If $\lambda$ is selected such that $\mathcal{A}^\lambda \approx 0$, then according to equation (\ref{eq:ep-tau}), one can obtain the estimates $\hat{\mathcal{B}}$ and $\hat{\mathcal{G}}$ by invoking the following minimization problem,
\begin{equation}\label{eq:sen-opt}
\begin{aligned}
& \underset{\mathcal{B}, \mathcal{G}_1}{\text{minimize}}
& & \|\mathcal{E}_{1,j}(k)-\mathcal{T}_{1,\lambda}\mathbf{\mathcal{Z}}_{(\lambda-1),j}(k-\lambda)\|_2 \\
& \text{subject to}
& & (\mathcal{B}(:,im+1:im+il))^{p+}=0.
\end{aligned}
\end{equation}
The constraint above does in fact enforce the columns $p,\ldots,ip$ of $\Delta_y$ to be equal to zero. For the case of sensor fault estimation problem, $\hat{A}_r$ is selected to be an arbitrary Hurwitz matrix, therefore $\hat{A}_r=A_r$. On the other hand,  $\Delta_y=\delta A=\hat{A}_r-\mathbf{M}+\hat{L}_r^{p\downarrow}={A}_r-\mathbf{M}+\hat{L}_r^{p\downarrow}=-{L}_r^{p\downarrow}+\hat{L}_r^{p\downarrow}$. Therefore, the columns $p,\ldots,ip$ corresponding to $\Delta_y$ should be equal to zero.\\}

{A methodology for solving the minimization problem (\ref{eq:sen-opt}) is provided in the Appendix \ref{app:opt-problem}. The solution will be consistent if  the matrix $\mathbf{\mathcal{Z}}_{(\lambda-1),j}(k)$ is full row rank which is not generally guaranteed. This condition on the matrix $\mathbf{\mathcal{Z}}_{(\lambda-1),j}(k)$ depends on the nature of the system feedback control, the excitation signal and the available data and the real model of the system (\cite{wan2015data,chiuso2007relation,van2013closed}).\\}

{If the above condition on $\mathbf{\mathcal{Z}}_{(\lambda-1),j}(k)$ is not satisfied, then the solution that is obtained by invoking the pseudo-inverse of $\mathbf{\mathcal{Z}}_{(\lambda-1),j}(k)$ still minimizes the cost function in (\ref{eq:sen-opt}), but it will be biased.\\}

{An estimate of the error dynamics in presence of the sensor faults is therefore given by,
\begin{equation}\label{eq:residual-error-dyn3}
\left\lbrace\begin{array}{l}
\hat{\xi}(k+1)=\hat{A}_r\hat{\xi}(k)+\hat{\mathcal{B}}\mathbf{Z}(k-i)-\hat{{\Delta}}_y\mathbf{F}^s(k-i)\\ \Delta \hat{f}^{s}(k)=\mathbf{I}_l^l\left(\hat{\xi}(k)+\hat{\mathcal{G}}_1\mathbf{U}(k-i)\right)\end{array}\right.
\end{equation}
where $\Delta \hat{f}^{s}(k)$ is an estimate of $\Delta {f}^{s}(k)$. We assumed that the sensors $\sim p$ are healthy for the purpose of fault estimation of the sensors $p$. Moreover,  the columns $p,\ldots,ip$ of $\hat{\Delta}_y$ are enforced to be zero in the minimization problem (\ref{eq:sen-opt}). Therefore, $\hat{{\Delta}}_y\mathbf{F}^s(k-i)$ is practically zero.\\}

{Consequently, one can now construct a new and a so-called \textit{tuned} sensor fault estimation filter that is governed by,
\begin{equation}\label{eq:sen-est-modif}
\left\lbrace\begin{array}{l}
\eta(k+1)=\hat{A}_r\eta(k)+\tilde{\mathcal{B}}_s\mathbf{Z}(k-i)\\ \hat{f}^{s}(k-i)=\mathbf{I}_{l}^l \left(\eta(k)-\mathbf{Y}(k-i)+\tilde{\mathcal{D}}_s\mathbf{U}(k-i)\right) 
\end{array}\right.
\end{equation}
where $ \hat{f}^{s}(k-i)$ denotes the tuned estimate of the sensor fault,  $\tilde{\mathcal{B}}_s=\left[\begin{array}{cc}\hat{B}_r &\hat{L}^{p\downarrow}\end{array}\right]+\hat{\mathcal{B}}$ and $\tilde{\mathcal{D}}_s=\hat{\mathbf{D}}+\hat{\mathcal{G}}_{1}$, where $\hat{\mathcal{B}}$ and $\hat{\mathcal{G}}_{1}$ are obtained and given by the minimization problem (\ref{eq:sen-opt}). \\}
%%%%%%%%%%%%%%%%%%%%%%%%%%%%%%%%%%%%%%%%%%%%%
{\subsection{Actuator Fault Estimation Filters}
The same procedure can now be followed for the actuator fault estimation filter. First, the following fault filter estimation is considered for the actuators $q$ by using the data from the sensors $\sim p$ and actuators $\sim q$, namely  
\begin{equation}\label{eq:actuator-est}
\left\lbrace\begin{array}{l}
\eta(k+1)=\hat{A}_r\eta(k)+\hat{B}_r\mathbf{U}^{q-}(k-i)+\hat{L}_r\mathbf{Y}^{p-}(k-i)\\ \doublehat{f}^{a}(k-i)=-\mathbf{I}_{m}^m\hat{\mathbf{D}}_{p-}^\dagger(\eta(k)-  \\
\mathbf{Y}^{p-}(k-i)+\hat{\mathbf{D}}_{p-}\mathbf{U}(k-i))
\end{array}\right.
\end{equation}
where $\doublehat{f}^{a}(k-i)$ denotes an estimate of $f^a(k-i)$ and the filter parameters are obtained from the solution to the equations (\ref{eq:eq1}) to (\ref{eq:eq4}).\\}

{However, following along the same lines as those used in the Subsection \ref{subsec:sen-est}, the so-called \textit{tuned} actuator fault estimation filter is now proposed as follows,
\begin{equation}\label{eq:act-est-modif}
\left\lbrace\begin{array}{l}
\eta(k+1)=\hat{A}_r\eta(k)+\tilde{\mathcal{B}}_a\mathbf{Z}^{q-,p-}(k-i)\\ \hat{f}^{a}(k-i)=-\mathbf{I}_{m}^m\tilde{\mathcal{D}}_a^\dagger(\eta(k)- \\ \mathbf{Y}^{p-}(k-i)+\tilde{\mathcal{D}}_a\mathbf{U}(k-i)) 
\end{array}\right.
\end{equation}
where $\tilde{\mathcal{B}}_a=\left[\begin{array}{cc}\hat{B}_r &\hat{L}_{r}\end{array}\right]+\hat{\mathcal{B}}$ and $\tilde{\mathcal{D}}_a=\hat{\mathbf{D}}_{p-}+\hat{\mathcal{G}}_1$. The parameters $\hat{\mathcal{B}}$ and $\hat{\mathcal{G}}_{1}$ are obtained by invoking the following minimization problem,
\begin{equation}\label{eq:act-opt}
\begin{aligned}
& \underset{\mathcal{B}, \mathcal{G}_1}{\text{minimize}}
& & \|\mathcal{E}_{1,j}(k)-\mathcal{T}_{1,\lambda}\mathbf{\mathcal{Z}}^{q-,p-}_{(\lambda-1),j}(k-\lambda)\|_2 
\end{aligned}
\end{equation}
where the matrices $\mathcal{E}_{1,j}(k)$ and $\mathbf{\mathcal{Z}}^{q-,p-}_{(\lambda-1),j}(k)$ are constructed similar to $\mathbf{G}_{i,j}(k)$ by replacing the signal $g(k)$ with the signals $-\doublehat{f}^{a}(k-i)$ and $\mathbf{Z}^{q-,p-}(k)$, respectively, and where $\mathbf{Z}^{q-,p-}(k)=\left[\begin{array}{c}\mathbf{U}^{q-}(k)\\\mathbf{Y}^{p-}(k)\end{array}\right]$, and $j$ is selected as large as the available data corresponding to the second segment allows. Note that the signal $-\doublehat{f}^{a}(k-i)$ for construction of the matrix $\mathcal{E}_{1,j}(k)$ is obtained by stimulating the filter (\ref{eq:actuator-est}) with the second segment of the healthy I/O data using an arbitrary initial condition for $\eta (0)$. \\}

{The above minimization problem is solved similar to the problem (\ref{eq:sen-opt}) as described in the Appendix \ref{app:opt-problem}. We will demonstrate in the next section that the above tuning procedures will significantly improve the faults estimation accuracy performance. Similarly, the same tuning procedure can be applied to the fault detection and isolation filters in order to improve their performance when applied to a specific application. However, these details are left as topics of our future work. }
{\begin{rem}
As stated earlier in this section, one should partition the available off-line I/O data before applying the above tuning procedure. The length of the data in the segment that is used for solving the minimization problem (\ref{eq:sen-opt}) or (\ref{eq:act-opt}) must be at least greater than $\lambda$, where $\lambda$ was selected such that $\mathcal{A}^\lambda \approx 0$. 
\end{rem}} 
%%%%%%%%%%%%%%%%%%%%%%%%%%%%%%%%%%%%%%%%%%%%%%%%%%%%%%%%%%%%%%%%%%%%%%%%%%%%%%%%%%%%%%%%%%%%%%%%%%
\section{Simulation Results}\label{sec:simulation}
In this section, we provide two numerical examples and simulations to illustrate the merits and advantages of our proposed schemes. In both cases, the healthy input is generated by a Pseudo Random Binary Signal (PRBS) generator. The system healthy output is generated by simulating it subject to healthy input in addition to state and measurement noise ($\mathcal{N}(0,0.1)$) as governed by the dynamics  $\mathbf{S}$. The Markov parameters are estimated by using the MATLAB built-in function \textit{impulseest}.

{\textbf{Fault Detection and Isolation Results}: We consider the following non-minimum phase system which includes the fault model for the actuator bias $(f^a_k)$ and sensor bias $(f^s_k)$ as additive terms,
\begin{eqnarray}\label{sim:sys1}
x_{k+1}&=&\left[\begin{array}{cccc}0&0&0&-0.01\\1&0&0&0.08\\0&1&0&-0.27\\0&0&1&-0.54\end{array}\right]x_k+\left[\begin{array}{cc}1 &-0.3\\0 &3.82\\0&1.55\\0&-0.61\end{array}\right](u_k+f^a_k) \nonumber \\
y_{k}&=&\left[\begin{array}{ccccc}1.58&0.725&-0.60&0.31\\2.4 &-0.08&0.42&-0.05 \end{array}\right]x_k+f^s_k
\end{eqnarray}
The poles and zeros of the above system are located at  $\{-0.39\pm53j,0.11\pm 0.09j\}$ and $\{0.17,1.49\}$, respectively. Figure \ref{fig:sys1}a shows the output of the residual generator filter (equation (\ref{eq:general-form})) for performing the fault detection task by setting $p=q=\{\emptyset\}$ when a bias fault is injected in the actuator 1 at the time instant $k=150$. We set $i=2$. The identification data include 1000 samples. The numerical values for the detection filter are provided in the Appendix \ref{app:detfilter}. Figures \ref{fig:sys1}b and \ref{fig:sys1}c depict the outputs of the fault isolation filters 1 and 2 having the setting $q=\{1\}, p=\{\emptyset\}$ and $q=\{2\}, p=\{\emptyset\}$, respectively, and $i=2$ for both. We have not yet applied the tuning process that was discussed in Section \ref{sec:estimation} to the above results, nevertheless these results demonstrate that actuator faults are successfully detected and isolated by application of our proposed data-driven methodology. In the next example, we will demonstrate the effects of the filter tuning process on the performance of the fault estimation accuracy.\\
\begin{figure}
  \centering
  \includegraphics[width=0.47\textwidth]{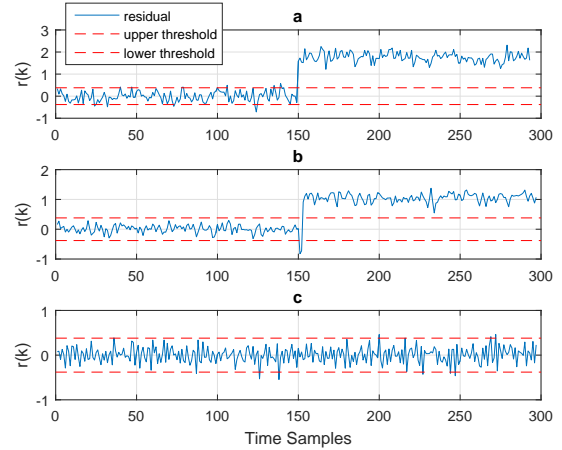}
%   \captionsetup{justification=centering}
  \caption{A fault is injected in the actuator 1 of the system (\ref{sim:sys1}) at $k=150$. (a) The output of the residual generator filter for achieving the fault detection task, (b) The output of the residual generator filter insensitive to the fault in the actuator 2, and (c) The output of the residual generator filter that is insensitive to the fault in the actuator 1.}\label{fig:sys1}
\end{figure}}

{
\textbf{Fault Estimation Results}: Consider now the following minimum phase system,
\begin{eqnarray}\label{sim:sys2}
x_{k+1}&=&\left[\begin{array}{cccc}-0.05&-0.40&0&-0.08\\-0.29&-0.11&0.05&-0.03\\-0.06&0.18&-0.43&0.36\\0.28&0.18&-0.43&0.36\end{array}\right]x_k \nonumber\\
&+&\left[\begin{array}{cc}-0.15 &-0.99\\0 &0\\-0.68&0.07\\-0.96&-0.20\end{array}\right](u_k+f^a_k) \nonumber \\
y_{k}&=&\left[\begin{array}{ccccc}-2.08&0&-0.69&0\\0 &-0.84&0.20&0.89 \end{array}\right]x_k+f^s_k
\end{eqnarray}
The poles and zeros of the system are located at  $\{-0.37,0.30,-0.51\pm 0.52j\}$ and $\{0.08,-0.58\}$, respectively. We next present a typical simulation result for estimating a fault in the system (\ref{sim:sys2}), and then provide comprehensive Monte Carlo simulations. Assume that a fault having a severity of 2 is injected in the sensor 2 at the time step $k=150$. We selected a relatively large amplitude input signal given below to \underline{magnify} the presence of biases,
\begin{equation}\label{eq:test-input}
u(k)=\left[\begin{array}{c} 20+20\sin(5k) \\ 30+30\cos(7k) \end{array}\right]
\end{equation}
We set $i=2$, $p=\{2\}$ and $q=\{\emptyset\}$. We used 700 data samples for estimation of the Markov parameters and $\hat{\mathbf{M}}$ (equation (\ref{eq:m-solution})) and 300 samples for the filter tuning process. First, we tested the performance of our proposed sensor fault estimator (\ref{eq:sensor-est}) that is shown in Figure \ref{fig:lpn0}(a). The results clearly indicate that the filter is biased with an estimation error of 24\%. The numerical values of the filter matrices are given in the Appendix \ref{app:lpn0}. We then tune the filter by solving the minimization problem (\ref{eq:sen-opt}) and construct the estimation filter as described by equation (\ref{eq:sen-est-modif}). \\}

{The numerical values for the matrices of the tuned filter are given in the Appendix \ref{app:lpn0}. The resulting estimation error for the tuned filter is now 1\% as shown in Figure \ref{fig:lpn0}(b) which illustrates a significantly improved and enhanced performance as compared to those depicted in Figure \ref{fig:lpn0}(a). A better illustration of the improved performance is now provided through Monte Carlo simulation runs as described below.
\begin{figure}
  \centering
  \includegraphics[width=0.47\textwidth]{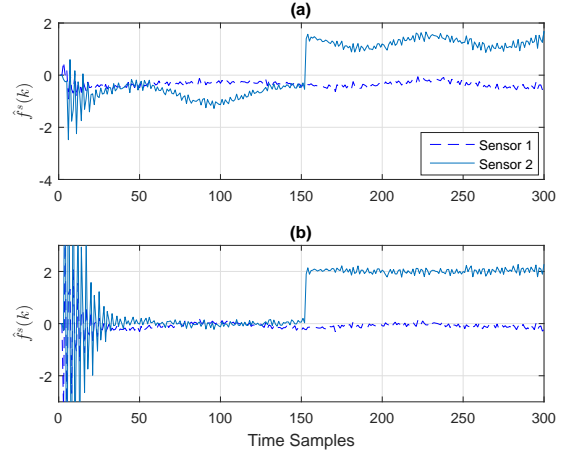}
%   \captionsetup{justification=centering}
  \caption{A fault having a severity of 2 is injected in the sensor 2 of the system (\ref{sim:sys2}) at $k=150$. (a) The output of the original sensor fault estimation filter, and (b) The output of the tuned sensor fault estimation filter.}\label{fig:lpn0}
\end{figure}
\begin{table}
\centering
\caption{The Monte Carlo simulation results for estimation of the faults in the system (\ref{sim:sys2}) using different filters and under two different system inputs, where $\mu$ and $\sigma$  denote the mean and variance, respectively. The filters are specifically the sensor fault estimator (\ref{eq:sensor-est}), the tuned sensor fault estimator (\ref{eq:sen-est-modif}), the actuator fault estimator (\ref{eq:actuator-est}) and the tuned fault estimator (\ref{eq:act-est-modif}) denoted by F(\ref{eq:sensor-est}), F(\ref{eq:sen-est-modif}), F(\ref{eq:actuator-est}) and F(\ref{eq:act-est-modif}), respectively. }
\label{tab:mont-carlo}
\begin{tabularx}{0.45\textwidth}{X|X|X|X|X|}
\cline{2-5}
                       & \multicolumn{2}{c|}{$u_1(k)$} & \multicolumn{2}{c|}{$u_2(k)$} \\ \cline{2-5} 
                       &      $\mu (\Delta f)$     &    $\sigma (\Delta f)$       &    $\mu (\Delta f)$       &     $\sigma (\Delta f)$      \\ \hline
\multicolumn{1}{|l|}{F(\ref{eq:sensor-est})} &   (-0.47, -0.06)        &     (0.30, 1.13)      &  (-0.03, 0)         &   (0.0, 0.02)        \\ \hline
\multicolumn{1}{|l|}{F(\ref{eq:sen-est-modif})} &   (-0.02, 0)        & (0.07, 0.18)          &     (-0.04, 0)      &     (0, 0.01)      \\ \hline
\multicolumn{1}{|l|}{F(\ref{eq:actuator-est})} &    (1.31, -4.02)       &   (3.2, 11.3)        &  (0.14, -0.4)         &    (0.02, 0.09)       \\ \hline
\multicolumn{1}{|l|}{F(\ref{eq:act-est-modif})} &    (-0.01, 0.01)       &    (0.01, 0.02)       &  (0, 0)         &    (0, 0)       \\ \hline
\end{tabularx}
\end{table} 
}
{\textbf{Monte Carlo Simulation Results}: We have conducted Monte Carlo simulation runs for estimation of the faults in the system (\ref{sim:sys2}). We set $i=2$, $p=\{1,2\}$ and $q=\{\emptyset\}$  and $i=2$, $p=\{\emptyset\}$ and $q=\{1,2\}$ for the sensor and the actuator fault estimations, respectively. Sensors faults having severities of -1 and 1 are injected to the sensor 1 and the sensor 2  at the time step $k=150$, respectively. The same fault scenario is considered for the actuator fault estimation problem. We performed 400 Monte Carlo simulation runs for two inputs that are selected as $u_1(k)=u(k)$ given by equation (\ref{eq:test-input}) and $u_2(k)=0.1u(k)$. The results are shown in Figure \ref{fig:mont1} and numerically presented in Table \ref{tab:mont-carlo}. It can be concluded that the filters (\ref{eq:sensor-est}) and (\ref{eq:actuator-est}) have an acceptable performance for relatively small inputs (in terms of the norm of the signal). On the other hand, relatively large inputs clearly magnify the biases although they are well-managed by utilizing our proposed tuning process. An approximation to the biases for the filters (\ref{eq:sensor-est}) and (\ref{eq:actuator-est}) can be obtained by using equation (\ref{eq:residual-error-dyn}). The $L_2$ gain of the error dynamics is then given by,
\begin{equation*}
\|\Delta f^{s}(k)\|_2 \leq \|(zI-\hat{A}_r)\hat{\mathcal{B}}+\hat{\mathcal{G}}\|_\infty\|\mathbf{Z}(k)\|_2
\end{equation*}
The matrices $\hat{\mathcal{B}}$ and $\hat{\mathcal{G}}_{1}$ are obtained by solving the minimization problems (\ref{eq:sen-opt}) and (\ref{eq:act-opt}). Therefore, one can obtain a prediction of the error margin corresponding to a certain input.
\begin{figure}
  \centering
  \includegraphics[width=0.45\textwidth]{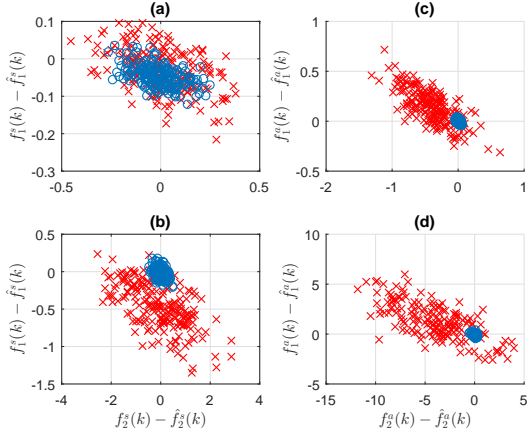}
%   \captionsetup{justification=centering}
  \caption{Faults having severities of 1 and -1 are injected in the actuators or sensors of the system (\ref{sim:sys2}). The red marks represent results for the filters (\ref{eq:sensor-est}) or (\ref{eq:actuator-est}) and the blue circles represent results for the tuned filters (\ref{eq:sen-est-modif}) or (\ref{eq:act-est-modif}). (a) Sensors fault estimation error when the system is stimulated by $u_2(k)$, (b) Sensors fault estimation error when the system is stimulated by $u_1(k)$, (c) Actuator fault estimation errors when the system is stimulated by $u_2(k)$, and (d) Actuator fault estimation errors when the system is stimulated by $u_1(k)$. }\label{fig:mont1}
\end{figure}
}

\textbf{Comparative Study:} Finally, in order to perform a \underline{comparative study} to demonstrate the capability and advantage of our proposed methodology, we consider the example that was provided in \cite{wan2015data} and evaluate our corresponding results with those in \cite{wan2015data}. The system in \cite{wan2015data} is a continuous-time system and represents a linearized model of a vertical take-off and landing (VTOL) aircraft that is given by,
% \begin{tiny}
  \begin{eqnarray}\label{sim:syscomp}
\dot{x}(t)&=&\left[\begin{array}{cccc}-0.036&0.027&0.018&-0.455\\0.048&-1.01&0.002&
-4.020\\0.100&0.368&-0.707&1.42\\0&0&1&0\end{array}\right]x(t)\nonumber \\
&+&\left[\begin{array}{cc}0.44 &0.17\\3.54 &-7.59\\-5.52&4.49\\0&0\end{array}\right](u(t)+f^a(t)) \nonumber \\
y(t)&=&\left[\begin{array}{cccc}1&0&0&0\\0&1&0&0 \\0&0&1&0 \\ 0&1&1&1\end{array}\right]x(t)+f_s(t)
\end{eqnarray}
% \end{tiny}
where $f^s(t)\in \mathbb{R}^4 $ and $f^a(t) \in \mathbb{R}^2$ and  $f^a(t)$ with $f^s(t)$  representing the actuator and sensor bias faults, respectively. The discrete-time model associated with the system (\ref{sim:syscomp}) is obtained by using a sampling rate of 0.5 seconds. Furthermore, it is assumed that the system is stabilized by applying the following control law which is experimentally obtained,
\[u(k)=-\left[\begin{array}{cccc}0&0&-0.5&0\\0&0&-0.1&-0.1 \end{array}\right]y(k)+\xi(k)\]
where $\xi(k)$ denotes the reference signal. The process and measurement noise are  white having zero mean and covariances $Q=0.16\mathbf{I}$ and $R=0.64\mathbf{I}$, respectively. The reference signal is selected to be a PRBS signal for identification of the Markov parameters. \\

The identification data set includes 1000 samples. {Note that the \textit{correlation analysis} cannot be directly applied to unstable systems. Therefore,  response of the stable closed-loop system is obtained by injecting the input that is computed at each time step using the above control law. Next, the input and closed-loop system responses are used as I/O data for identification of the closed-loop system Markov parameters through the \textit{correlation analysis}}. The injected fault signals to the actuators and sensors are given by,
\[f^a(k)=\left\lbrace\begin{array}{lr} \left[\begin{array}{cc}0&0\end{array}\right]^T& 0\leq k \leq 50 \\ \left[\begin{array}{cc}sin(0.1\pi k)&1\end{array}\right]^T&k>50\end{array}\right.\]
\[f^s(k)=\left\lbrace\begin{array}{lr} \left[\begin{array}{cccc}0&0&0&0\end{array}\right]^T& 0\leq k \leq 50 \\ \left[\begin{array}{cccc}sin(0.1\pi k)&1&0&0\end{array}\right]^T&k>50\end{array}\right.\]
We have set $i=2$, $p=\{1,2\}$ and $q=\{\emptyset\}$ and $i=3$, $p=\{\emptyset\}$ and $q=\{1,2\}$ for the sensor and actuator fault estimation filters, respectively. The reference signal is set to $\xi(k)=15$ in order to duplicate the Monte Carlo simulation results that were reported in \cite{wan2015data}.  

\begin{figure}
  \centering
  \begin{tabular}{cc}
    a  \\
    \includegraphics[width=0.45\textwidth]{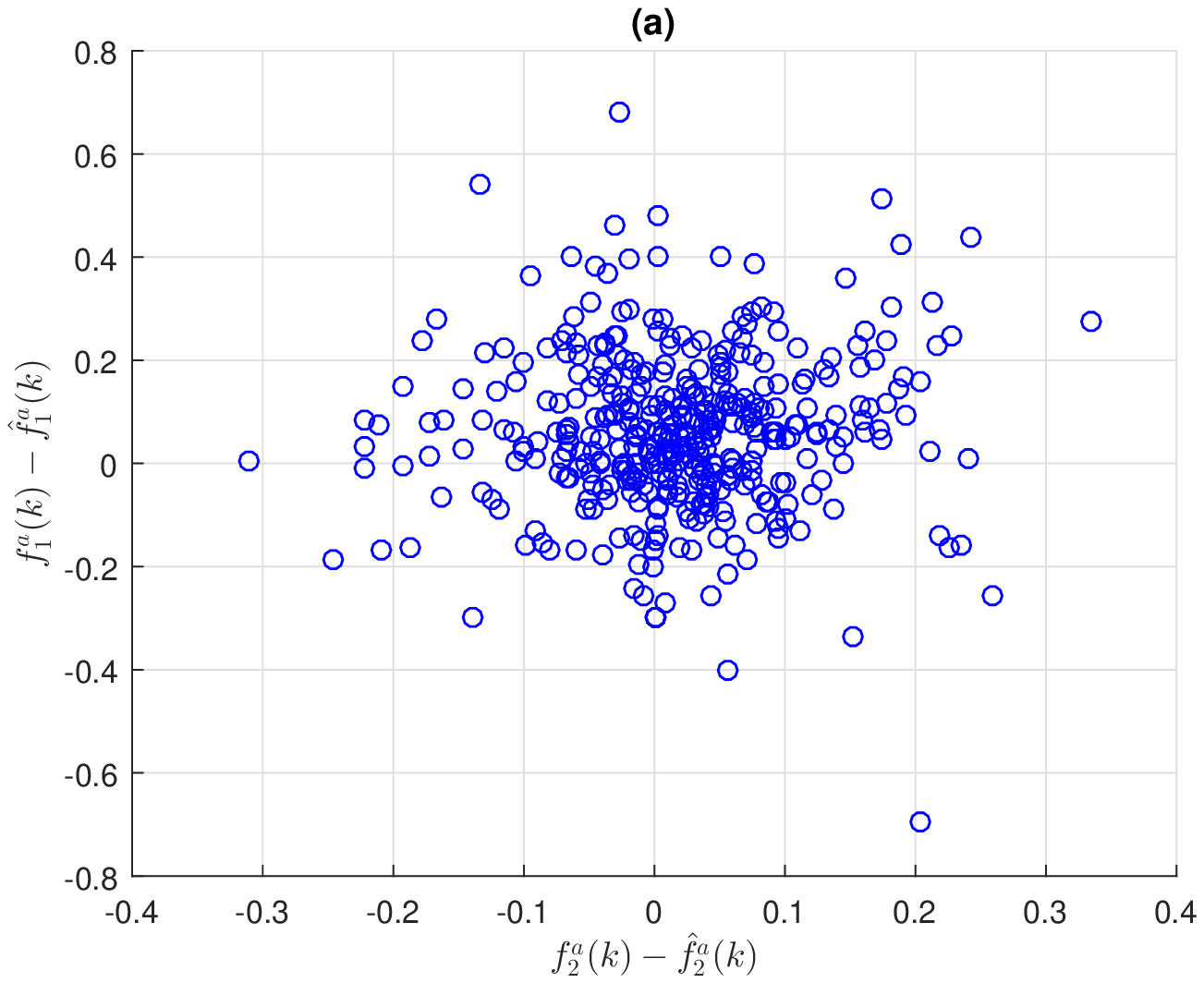}
      \\b \\ \includegraphics[width=0.45\textwidth]{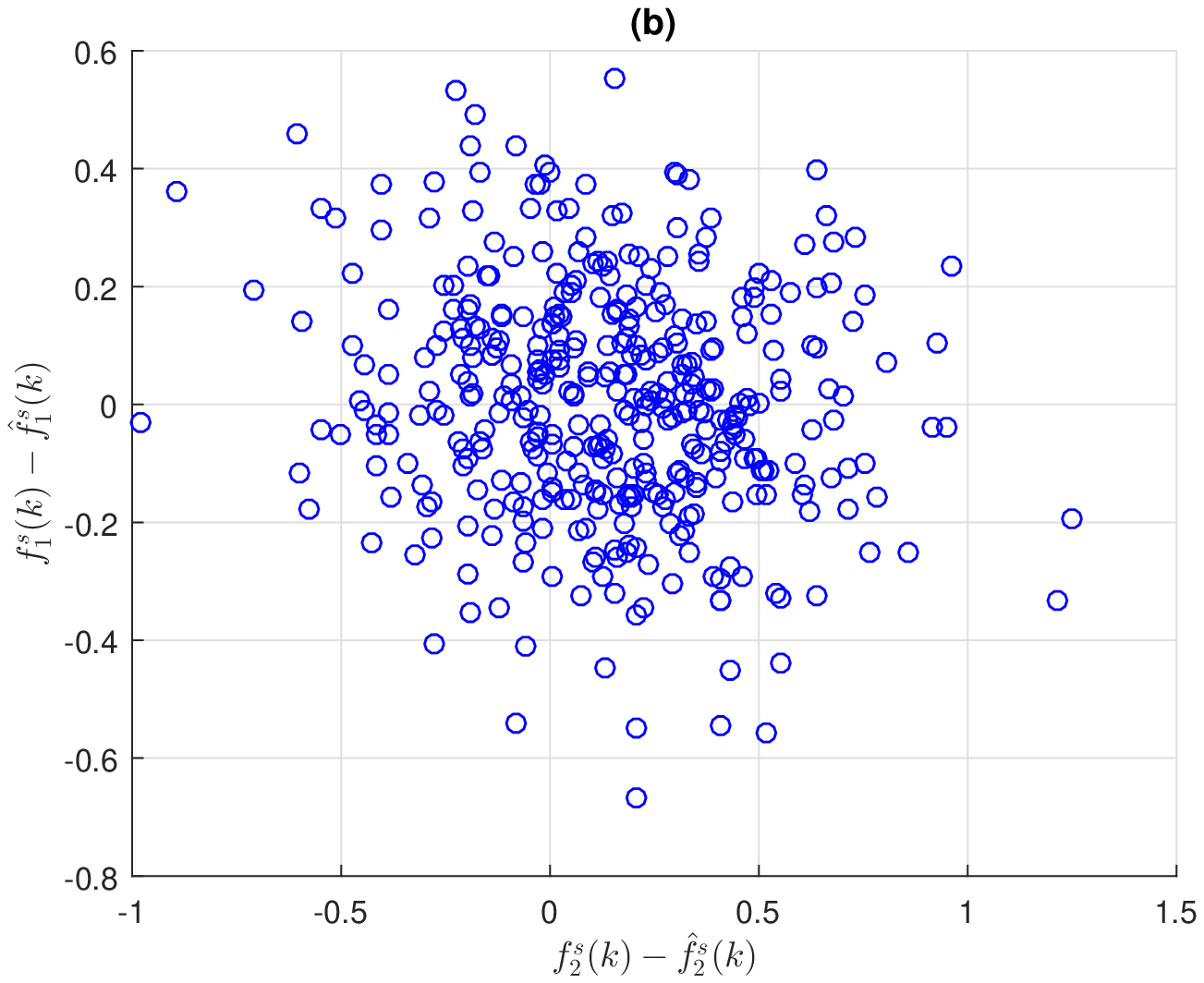} \\
  \end{tabular}
%   \captionsetup{justification=centering}
    \caption{(a) The first actuator fault estimation error versus the second actuator fault estimation error for the system (\ref{sim:syscomp}) using 500 Monte Carlo simulation runs, and (b) The first sensor fault estimation error versus the second sensor fault estimation error for the system (\ref{sim:syscomp}) using 400 Monte Carlo simulation runs. }\label{fig:comp}
\end{figure}

Figures \ref{fig:comp}a and \ref{fig:comp}b show 500 and 400 Monte Carlo simulation runs for estimation of the actuator and sensor faults, respectively. The average estimation errors are given by $\mu(\hat{f}_1^a-f_1^a, \hat{f}_2^a-f_2^a)=(0.018,0.039)$ and $\mu(\hat{f}_1^s-f_1^s, \hat{f}_2^s-f_2^s)=(0.005,0.139)$. The variances are given by $\sigma(\hat{f}_1^a-f_1^a, \hat{f}_2^a-f_2^a)=(0.008,0.0097)$ and $\sigma(\hat{f}_1^s-f_1^s, \hat{f}_2^s-f_2^s)=(0.0398,0.1074)$.

{The above results clearly show that our proposed scheme has significant advantages, benefits, and capabilities over the receding horizon fault estimator that was proposed in \cite{wan2015data}, although it uses the same set of assumptions. This is substantiated by the following observations. First, our proposed filter order is significantly lower than that in \cite{wan2015data} as we have theoretically shown in Theorems \ref{thm:actuator-bias-org} and \ref{thm:sensor-bias-org}. For this particular example, we have used $i=2$ for the sensor and actuator fault estimation filters, respectively, whereas $i$ is set to $i=30$ in \cite{wan2015data}. Moreover, we have achieved a better performance by invoking an offline tuning procedure as compared to the Algorithm 3 utilized in \cite{wan2015data} that performs an online optimization solution. Consequently,  the computational burden of \cite{wan2015data} to the user increases to the point where the average computational time per sample takes 2.05 seconds on a 3.4 GHz computer having 8 GB of RAM. Whereas, the computational time associated with our proposed methodology per sample using the same computer takes only $8.2 \times 10^{-7}$ seconds.}

% \begin{figure}
%   \centering
%   % Requires \usepackage{graphicx}
%   \includegraphics[width=0.6\textwidth]{p3f5.pdf}\\
%   \caption{Actuator fault estimation errors for system (\ref{sim:sys2}) versus each other obtained from 500 simulations. Simulation set up is injection of faults equal to 2 and 1 in actuators 1 and 2, respectively. }\label{fig:mont-carlo}
% \end{figure}

%%%%%%%%%%%%%%%%%%%%%%%%%%%%%%%%%%%%%%%%%%%%%%%%%%%%%%%%%%%%%%%%%%%%%%%%%%%%%%%%%%%%%
\section{Conclusion}

We have proposed fault detection, isolation and estimation schemes that are all directly constructed and designed in the state-space representation form by utilizing only the system I/O data. We have shown that to design and develop our schemes it is only sufficient to estimate the system Markov parameters. Consequently, the reduction step that is commonly used in the literature, and that also introduces nonlinear errors, and also requires an \textit{a priori} knowledge of the system order is completely eliminated in our schemes. We have shown that the performance of the estimation scheme is linearly dependent on the Markov parameters estimation process errors. We also proposed an offline tuning procedure that effectively compensates for the estimation errors that are caused by errors in the estimation of the Markov parameters. Comparisons of our proposed schemes with those available in the literature have revealed that our methodology is mathematically simpler to develop and computationally more efficient, while it maintains the same level of performance and requires a lower set of assumptions. Further research is required to investigate the robustness of our scheme to estimation errors and presence of concurrent sensor and actuator faults.

\bibliographystyle{ieeetr}
\bibliography{engine}

%%%%%%%%%%%%%%%%%%%%%%%%%%%%%%%%%%%%%%%%%%%%%%%%%%%%%%%%%%%%%%%%%%%%%%%%%%%%%%%%%%%%%%%%%%%%%%
\appendix
{
\section{Proof of Theorem \ref{thm:actuator-bias-org} }\label{app:actuator-bias-org}
The dynamics of the residuals in presence of actuator faults is given by,
\begin{equation}
\left\lbrace\begin{array}{l}
\eta(k+1)={A}_r\eta(k)+{B}_r^{q-}(\mathbf{U}^{q-}(k-i)+\mathbf{F}^{a,q-}(k-i)\\{L}_r^{p-}\mathbf{Y}^{p-}(k-i)\\ {r}_q(k)=\eta(k)-\mathbf{Y}^{p-}(k-i)\\ +{\mathbf{D}}_{p-}(\mathbf{U}(k-i)+\mathbf{F}^a(k-i))
\end{array}\right.
\end{equation}
Since $A_r$, $B_r$ and $L_r$ satisfy equations (\ref{eq:M-definition-org}) and (\ref{eq:eq1org}) to (\ref{eq:eq4org}) and the actuators $\sim q$ are healthy, therefore $\mathbb{E}\{\eta(k) - \mathbf{C}_{p-}x(k-i)\} \rightarrow 0$ as $k \rightarrow \infty$. The convergence is asymptotic since $A_r$ is Hurwitz. Therefore, $\mathbb{E}\{r_q(k)\} \rightarrow -\mathbf{D}_{p-} \mathbf{F}^a(k-i)$ as $k \rightarrow \infty$. Therefore, $\hat{f}^a(k-i)=-\mathbf{I}_{l'}^m\mathbf{D}^\dagger\mathbb{E}\{r_q(k)\}-\mathbf{I}_{m}^m(\mathbf{I}-\mathbf{D}_{p-}^\dagger \mathbf{D}_{p-})\Theta$, where $\Theta$ is an arbitrary matrix. However, since $i \geq \tau_p$, the subspace spanned by rows of $\mathbf{I}_{m}^m$ are also spanned by the rows of $\mathbf{D}_{p-}$. Therefore, the projection of the row space of $\mathbf{I}_{m}^m$ onto the null space of $\mathbf{D}_{p-}$ given by $(\mathbf{I}-\mathbf{D}_{p-}^\dagger \mathbf{D}_{p-})$ is zero. In other words, $\mathbf{I}_{m}^m(\mathbf{I}-\mathbf{D}_{p-}^\dagger \mathbf{D}_{p-})=0$. This completes the proof of the theorem.
%%%%%%%%%%%%%%%%%%%%%%%%%%%%%%%%%%%%%%%%%%%%%%%%%%%%%%%%%%%%%%%%%
\section{Proof of Theorem \ref{thm:sensor-bias-org}}\label{app:sensor-bias-org}
The dynamics of the residuals in presence of sensor faults is given by,
\begin{equation}
\left\lbrace\begin{array}{l}
\eta(k+1)={A}_r\eta(k)+{B}_r\mathbf{U}(k-i)\\ +{L}_r^{p\downarrow}(\mathbf{Y}(k-i)+\mathbf{F}^s(k-i)\\ {r}_p(k)=\eta(k)-\mathbf{Y}(k-i)-\mathbf{F}^s(k-i)+{\mathbf{D}}_{p-}\mathbf{U}(k-i)
\end{array}\right.
\end{equation}
Since $A_r$, $B_r$ and $L_r^{p \downarrow}$ satisfy equations (\ref{eq:M-definition-org}) and (\ref{eq:eq1org}) to (\ref{eq:eq4org}) and the actuators and sensors $~p$ are healthy, therefore $\mathbb{E}\{\eta(k) - \mathbf{C}_{p-}x(k-i)\} \rightarrow 0$ as $k \rightarrow \infty$. The convergence is asymptotic since $A_r$ is Hurwitz. Therefore, $\mathbb{E}\{r_p(k)\} \rightarrow  \mathbf{F}^s(k-i)$ as $k \rightarrow \infty$. Therefore, $\hat{f}^s(k-i)=\mathbf{I}_{l}^m\mathbb{E}\{r_p(k)\}$. This completes the proof of the theorem.
%%%%%%%%%%%%%%%%%%%%%%%%%%%%%%%%%%%%%%%%%%%%%%%%%%%%%%%%%%%%%%%%%%%%%%
\section{Proof of Lemma \ref{lem:m-consitent}}\label{app:gamma-rank}
According to equation (\ref{eq:gamma0}) and the measurement equation (\ref{eq:system-subspace}), we have 
\[\Gamma_0^{p-}=\mathbf{C}_{p-}X(k-i)+\mathbf{E}_{p-}\mathbf{W}_{i,j}(k-i)+\mathbf{V}_{i,j}^{p-}(k-i)\]
where,
\[X(k-i)=\left[\begin{array}{ccc}x(k-i)& \ldots&x(k-i+j-1)\end{array}\right]\]
Assume that two block rows $\gamma$ and $\beta$ of $\Gamma_0^{p-}$, where $\beta > \gamma$, are linearly dependent which implies,
% \begin{tiny}
\begin{multline}
\mathbf{Y}_{1,j}^{p-}(k-i+\gamma)\\-(\mathbf{D}_{p-}((i+\gamma-1)l':(i+\gamma)l',:))\mathbf{U}_{i,j}(k-i)\\
=c(\mathbf{Y}_{1,j}^{p-}(k-i+\beta)\\ -(\mathbf{D}_{p-}((i+\beta-1)l':(i+\beta)l',:))\mathbf{U}_{i,j}(k-i))
\end{multline}
% \end{tiny}
where $c$ is a constant. Equivalently, we have,
% \begin{tiny}
\begin{multline}\label{eq:X-E-V}
CA^{\gamma-1}X(k-i)\\+(\mathbf{E}_{p-}((i+\gamma-1)l':(i+\gamma)l',:))\mathbf{W}_{i,j}(k-i)\\+\mathbf{V}_{1,j}^{p-}(k-i+\gamma)=c(CA^{\beta-1}X(k-i)\\+(\mathbf{E}_{p-}((i+\beta-1)l':(i+\beta)l',:))\mathbf{W}_{i,j}(k-i)+\mathbf{V}_{1,j}^{p-}(k-i+\beta))
\end{multline}
% \end{tiny}
If we multiply both sides of equation (\ref{eq:X-E-V}) by $(\mathbf{V}_{1,j}^{p-}(k-i+\beta))^T$ and take the limit as $j \rightarrow \infty$, all the terms will be zero except for the last one since all the terms except the last one are uncorrelated with $(\mathbf{V}_{1,j}^{p-}(k-i+\beta))$. Therefore, we obtain $0=c$, which is a contradiction. Therefore, $\hat{\Gamma}_0^{p-}$ is full row rank.
%%%%%%%%%%%%%%%%%%%%%%%%%%%%%%%%%%%%%%%%%%%%%%%%%%%%%%%%%%%%%%%
\section{Proof of Lemma \ref{lem:mhat-hurwitz}}\label{app:m-hurt}
Let us define $\psi(k-i)=\mathbf{Y}_{i,j}^{p-}(k-i)-\hat{\mathbf{D}}_{p-}\mathbf{U}_{i,j}(k-i)$. Threfore, $\psi (k-i+1)$ is governed by,
\begin{eqnarray}
\psi(k-i+1)&=&\mathbf{Y}_{i,j}^{p-}(k-i+1)-\hat{\mathbf{D}}_{p-}\mathbf{U}_{i,j}(k-i+1) \nonumber \\
&=&\mathbf{Y}_{(i+1),j}^{p-}(k-i+1)-\hat{\mathbf{D}}_{+,p-}\mathbf{U}_{(i+1),j}(k-i) \nonumber\\
&-&\hat{\mathcal{D}}_{+,p-}\mathbf{I}_{l'}^m\mathbf{U}_{i,j}(k-i)\nonumber \\
&=&\hat{\mathbf{M}}_{p-}\psi(k-i)-\hat{\mathcal{D}}_{+,p-}\mathbf{I}_{l'}^m\mathbf{U}_{i,j}(k-i)
\end{eqnarray}
Due to the fact that $\psi(k-i)$ is bounded, therefore $\hat{\mathbf{M}}_{p-}$ is a Hurwitz matrix. This completes the proof of the lemma.
%%%%%%%%%%%%%%%%%%%%%%%%%%%%%%%%%%%%%%%%%%%
\section{Proof of Lemma \ref{lem:trivial-zero}}\label{app:trivialZero}
We show that the first $(i-1)l$ rows of the residuals generated by the filter (\ref{eq:general-form}) approach to zero as $k \rightarrow \infty$ if $q=\{\emptyset\}$. We begin by noting that,
\begin{equation}
\hat{r}(k+1)=\eta(k+1)-\mathbf{Y}^{p-}(k-i+1)+\hat{\mathbf{D}}_{p-}\mathbf{U}(k-i+1)
\end{equation}
Substituting $\eta(k+1)$ from the state equation of the filter (\ref{eq:general-form}) yields,
\begin{multline}
\hat{r}(k+1)=\hat{A}_r\eta(k)+\hat{B}_r\mathbf{U}(k-i)+\hat{L}_{r}\mathbf{Y}(k-i)\\
-\mathbf{Y}^{p-}(k-i+1)+\hat{\mathbf{D}}_{p-}\mathbf{U}(k-i+1)
\end{multline}
Next, we substitute $\hat{B}_r$ and $\hat{L}_r$ from equations (\ref{eq:eq2}), (\ref{eq:eq3}) and (\ref{eq:eq4}) and rearrange it to obtain,
\begin{multline}
\hat{r}(k+1)=\hat{A}_rr(k)-\hat{\mathbf{M}}_{p-}\left(\mathbf{Y}^{p-}(k-i)-\hat{\mathbf{D}}_{p-}\mathbf{U}(k-i)\right)\\ -\mathbf{Y}^{p-}(k-i+1)+\hat{\mathbf{D}}_{p-}\mathbf{U}(k-i+1)\\ +\hat{\mathcal{D}}_{+,p-}\mathbf{I}_{l'}^m\mathbf{U}(k-i)
\end{multline}
Note that,
\[\hat{\mathbf{D}}_{p-}\mathbf{U}(k-i+1)+\hat{\mathcal{D}}_{+,p-}\mathbf{I}_{l'}^m\mathbf{U}(k-i)=\hat{\mathbf{D}}_{+,p-}\mathbf{U}_+(k-i)\]
Therefore, by considering the structure of $\hat{\mathbf{M}}_{p-}$ in Lemma \ref{lem:mhat-structure}, one can verify that (E.3) becomes,
\begin{equation}
\hat{r}(k+1)=\doublehat{A}_r \hat{r}(k)+\left[\begin{array}{c} 0_{(i-1)l \times 1} \\ y^{p-}(k+1)-\bar{\mathbf{D}}_{+,p-}\mathbf{U}_{+}(k-i)\end{array}\right]
\end{equation}
which shows that if $\hat{A}_r$ is a diagonal Hurwitz matrix, then the first  $(i-1)l$ rows of $\hat{r}(k)$ approach to zero as $k \rightarrow \infty$. Note that we did not use the relation $\hat{\mathbf{M}}\hat{\Gamma}_0(:,1)=\hat{\Gamma}_1(:,1)$ that only holds for the healthy system. In that case, we would clearly obtain $\hat{r}(k+1)=\doublehat{A}_r \hat{r}(k)+0$, which is a valid model. This completes the proof of the lemma.
%%%%%%%%%%%%%%%%%%%%%%%%%%%%%%%%%%%%%%%%%%%%%%%%%%%%%%%%%
\section{A Methodology for Solving the Minimization Problems (\ref{eq:sen-opt}) and (\ref{eq:act-opt})}\label{app:opt-problem}
% Recall that $\mathbf{\mathcal{Z}}^{q-,p-}_{(\lambda-1),j}(k)$ is constructed similar to $\mathbf{G}_{i,j}(k)$ by using the signal $\mathbf{Z}^{q-,p-}(k)$, where, $\mathbf{Z}^{q-,p-}(k)=\left[\begin{array}{c}\mathbf{U}^{q-}(k)\\\mathbf{Y}^{p-}(k)\end{array}\right]$ and $j$ is selected as large as the available data allows. Moreover, if we define $\mathcal{H}_\beta^{q-,p-}=\mathcal{C}_{p-}\mathcal{A}^\beta\mathcal{B}^{q-}$, then $\mathcal{T}_{1,\lambda}^{q-,p-}$ is given by,
% \begin{multline} \label{eq:dsi-form2}
% \mathcal{T}_{1,\lambda}^{q-,p-}=\\ \left(\begin{array}{cccccc} \mathcal{H}_{\lambda-1}^{q-,p-}&H_{\lambda-2}^{q-,p-}& \ldots & \mathcal{H}_{0}^{q-,p-} & \mathcal{G} \end{array} \right)
% \end{multline}
The constraint in the optimization problem enforces that columns that the multiplied by the rows $p, \ldots, ip$ of $\mathbf{Y}(k)$ should be zero. Therefore,  they can be simply removed by using $\mathcal{Z}^{\{\emptyset\},p-}(k)$ instead of $\mathcal{Z}(k)$ and by invoking now the following optimization problem,
\begin{equation}\label{eq:phi-opt}
\begin{aligned}
& \underset{\mathcal{B}^{\phi}, \mathcal{G}_1}{\text{minimize}}
& & \|\mathcal{E}_{1,j}(k)-\mathcal{T}^{\phi}_{1,\lambda}\mathbf{\mathcal{Z}}_{(\lambda-1),j}^{\{\emptyset\},p-}(k-\lambda)\|_2 
\end{aligned}
\end{equation}
where 
\begin{multline} \label{eq:dsi-form2}
\mathcal{T}_{1,\lambda}^{\phi}=\\ \left(\begin{array}{cccccc} \mathcal{C}\mathcal{A}^{\lambda-1}\mathcal{B}^\phi&\mathcal{C}\mathcal{A}^{\lambda-2}\mathcal{B}^\phi& \ldots & \mathcal{C}\mathcal{B}^\phi & \mathcal{G} \end{array} \right)
\end{multline}
Once $\hat{\mathcal{B}}^{\phi}$ is computed, $\hat{\mathcal{B}}$ is easily constructed by inserting back zero columns at the columns $im+p, im+2p, \ldots, im+ip$ of the matrix $\hat{\mathcal{B}}^{\phi}$. An estimate of $\hat{\mathcal{T}}_{1,\lambda}^{\phi}$ is now given by,
\begin{eqnarray}\label{eq:tau-sol}
\hat{\mathcal{T}}_{1,\lambda}^{\phi}&=&\mathcal{E}_{1,j}(k)\left(\mathbf{\mathcal{Z}}_{(\lambda-1),j}^{\{\emptyset\},p-}(k-s)\right)^\dagger \nonumber \\
&=&\left[\begin{array}{ccc}\hat{\mathcal{T}}_1 & \ldots &\hat{\mathcal{T}}_\lambda\end{array}\right]
\end{eqnarray}
where $\hat{\mathcal{T}}_\alpha \in \mathbb{R}^{l \times (im+il-in_p)}$. We reformulate the definition (\ref{eq:dsi-form2}) in the matrix form as follows,
\begin{equation}\label{eq:g-enforce}
\left[\begin{array}{c}\hat{\mathcal{T}}_1 \\ \vdots \\ \hat{\mathcal{T}}_\lambda\end{array}\right]=\left[\begin{array}{cc}\mathfrak{C} &0\\0 &I\end{array}\right]\left[\begin{array}{c}\hat{\mathcal{B}}^{\phi} \\ \hat{\mathcal{G}}_1\end{array}\right]
\end{equation}
 where,
 \begin{equation}
\mathfrak{C}=\left[\begin{array}{c}\mathcal{C}\\\mathcal{C}\mathcal{A} \\ \vdots \\ \mathcal{C}\mathcal{A}^{\lambda-1}\end{array}\right]
\end{equation}
Note that we have enforced $\mathcal{G}=\left[\begin{array}{cc}\mathcal{G}_1 &0\end{array}\right]$ in the right hand side of (\ref{eq:g-enforce}). The solution to (\ref{eq:g-enforce}) is now given by,
\begin{equation}
\left[\begin{array}{c}\hat{\mathcal{B}}^{\phi} \\ \hat{\mathcal{G}}_1\end{array}\right]=\left[\begin{array}{cc}\mathfrak{C} &0\\0 &I\end{array}\right]^\dagger \left[\begin{array}{c}\hat{\mathcal{T}}_1 \\ \vdots \\ \hat{\mathcal{T}}_\lambda\end{array}\right]
\end{equation}
Note that the matrix $\left[\begin{array}{cc}\mathfrak{C} &0\\0 &I\end{array}\right]$ is full column rank. The above solution provides the least square solution to the minimization problem (\ref{eq:phi-opt}). However, if $\mathbf{\mathcal{Z}}_{(s-1),j}^{\{\emptyset\},p-}(k-s)$ is full row rank, then the minimum value of zero will be achieved. For the problem (\ref{eq:act-opt}), it is only sufficient to replace $\mathbf{\mathcal{Z}}_{(\lambda-1),j}^{\{\emptyset\},p-}(k-\lambda)$, $\mathcal{T}_{1,\lambda}^{\phi}$ and  $\hat{\mathcal{B}}^{\phi}$ in equations  (\ref{eq:tau-sol}) and (\ref{eq:phi-opt}) by $\mathbf{\mathcal{Z}}^{q-,p-}_{(\lambda-1),j}(k)$, $\mathcal{T}_{1,\lambda}$, and $\hat{\mathcal{B}}$, respectively. This provides the details on the methodology for solving the minimization problems (\ref{eq:sen-opt}) and (\ref{eq:act-opt}).
%%%%%%%%%%%%%%%%%%%%%%%%%%%%%%%%%%%%%%%%%%%%%%%
\section{Numerical Values of the Matrices in Example 1}\label{app:detfilter}
\[\hat{\mathbf{M}}=\left[\begin{array}{cccc} 0.00 &    0.00&    1.00&    0.00\\
   0.00&   0.00&    0.00&    1.00\\
   -0.30&   0.12&   -0.51&    0.51\\
    0.28& -0.11&    0.31&  -0.16
\end{array}\right]\]
\[\hat{\bar{A}}_r=\left[\begin{array}{cc}0.26&0\\0 &0.44\end{array}\right];\hat{\bar{B}}_r=\left[\begin{array}{cccc} 0.70&  -0.99&         0&         0\\
    0.88 &   1.01 &        0  &       0
\end{array}\right]\]
\[\hat{\bar{L}}_r=\left[\begin{array}{cccc} -0.30&    0.12&  -0.78&    0.5159\\
    0.28&  -0.11&   0.31&  -0.61
\end{array}\right]\]
%%%%%%%%%%%%%%%%%%%%%%%%%%%%%%%%%%%%%%%%%%%%%%%%%%%%%%%%%%%
\section{Numerical Values of the Matrices in Example 2}\label{app:lpn0}
\[\hat{\mathbf{M}}=\left[\begin{array}{cccc} 0&   0&    1&   0\\
    0&   0&  0&    1\\
    0.06&   -0.18&   -0.19&   -0.17\\
   -0.17&  -0.43&  -0.48&   -0.81
\end{array}\right]\]
\[\hat{B}_r=\left[\begin{array}{cccc}   0.61&   1.61&         0   &      0\\
   -1.06&  -0.34&        0      &   0\\
   -0.28&  -0.51&      0      &   0\\
   -0.15& -1.71&        0      &   0
\end{array}\right];\hat{L}_r^{p\downarrow}=\left[\begin{array}{cccc}   0.12&        0 &   0.20&         0\\
    0.59&       0   & 0.08&         0\\
    0.24&        0  &  0.19&         0\\
    0.28&    0  &  0.59&        0
\end{array}\right]\]
\[\hat{A}_r=\hat{\mathbf{M}}-\hat{L}_r^{p\downarrow};\hat{\mathbf{D}}=\left[\begin{array}{cccc} 
         0    &     0  &       0  &       0\\
         0     &    0     &    0     &    0\\
    0.76  &  2.03 &         0   &      0\\
    -0.99 &  -0.16 &        0     &    0
\end{array}\right]\]
\[\hat{\mathcal{B}}(:,1:4)=\left[\begin{array}{cccc} 
         -0.07&   -0.03&    0.07&    0.14\\
    0.01&  -0.16&   0.03&   0.09\\
   0 -0.09&    0.03&   0.06 &0\\
   -0.04&   -0.16&   0&    0
\end{array}\right]\]
\[\hat{\mathcal{B}}(:,5:8)=\left[\begin{array}{cccc} 
 0.02&         0 &   0&         0\\
   -0.01&        0  &  0.02&        0\\
   -0.01&         0  &  0.08&         0\\
   -0.02&         0 &   0.17&        0
\end{array}\right];\hat{\mathcal{G}}=\left[\begin{array}{cccc} 
         0    &     0  &       0  &       0\\
         0     &    0     &    0     &    0\\
    -0.05  &  -0.07 &         0   &      0\\
   0 &  0 &        0     &    0
\end{array}\right]\]}
%%%%%%%%%%%%%%%%%%%%%%%%%%%%%%%%%%%%%%%%
\end{document}